\documentclass[journal=jctcce,manuscript=article]{achemso}

\usepackage[version=3]{mhchem} 

\usepackage{physics}
\usepackage{algorithm}
\usepackage{algpseudocode}
\usepackage[
    colorlinks=true, 
    linkcolor=blue, 
    filecolor=blue, 
    citecolor=blue, 
]{hyperref}
\usepackage{amsmath}
\usepackage{amssymb}
\usepackage{braket}
\usepackage{threeparttable} 
\usepackage{xcolor}

\newif\ifhighlight
 \highlightfalse

\definecolor{usccardinal}{rgb}{0.6, 0.0, 0.0}
\newcommand{\hy}[1]{\ifhighlight\textcolor{usccardinal}{#1}\else{#1}\fi}
\author{Wan Nie}
\affiliation[CityUHK]
{Department of Computer Science, City University of Hong Kong, Hong Kong SAR 999077, China}

\author{Songwei Liu}
\affiliation[CUHK]
{Department of Electronic Engineering, The Chinese University of Hong Kong, Hong Kong SAR 999077, China}

\author{Yingying Yu}
\affiliation[CityUHK]
{Department of Computer Science, City University of Hong Kong, Hong Kong SAR 999077, China}

\author{Zhiwen Wang}
\affiliation[CityUHK]
{Department of Computer Science, City University of Hong Kong, Hong Kong SAR 999077, China}

\author{Jun Yang}
\email{juny@hku.hk}
\affiliation[HKU]
{Department of Chemistry and HKU-CAS Joint Laboratory on New Materials, The University of Hong Kong, Hong Kong SAR 999077, China}
\alsoaffiliation[QAILab]
{Hong Kong Quantum AI Lab Limited, Hong Kong SAR 999077, China}

\title[An \textsf{achemso} demo]
  {Learning to Rank for Selected Configuration Interaction}

\abbreviations{IR,NMR,UV}

\begin{document}

\begin{tocentry}

\centering
\includegraphics[width=1.0\textwidth]{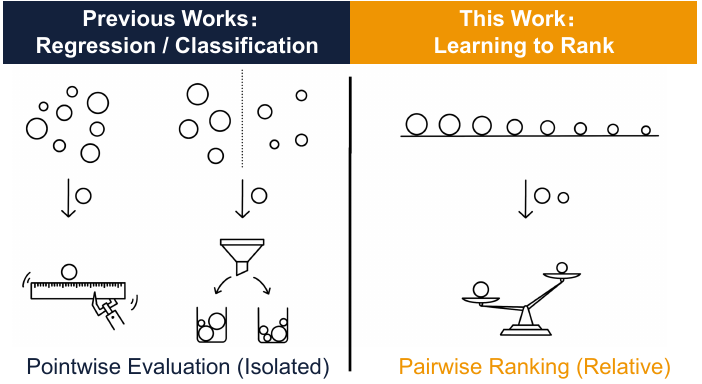}

\end{tocentry}

\begin{abstract}
The accurate description of electron correlation is a central challenge in computational chemistry, with selected configuration interaction (SCI) emerging as a powerful tool to approach the full CI limit. While recent machine learning (ML) integrations have accelerated determinant selection, existing regression and classification approaches suffer from a fundamental objective-loss mismatch: they evaluate the importance of determinants in isolation without explicitly accounting for their relative importance ranking. Here, we introduce ranking configuration interaction (RCI), a novel ML-supported SCI framework that reframes determinant selection as a pairwise ranking problem. Building upon a Transformer-based architecture to capture complex, non-local orbital dependencies, RCI progressively optimizes the partial ordering of determinants. By doing so, RCI aligns the training objective more closely with the intrinsic ranking nature of SCI. Extensive benchmarks across both plane-wave and Gaussian basis sets, including the molecules N$_2$, CO, H$_2$O, NH$_3$, and C$_2$, demonstrate the efficiency of RCI. Compared to previously reported classification baselines, RCI consistently accelerates convergence—reducing overall computational time by 23\% to over 50\% depending on the system, and requiring only 55\% of the determinant count in representative cases such as N$_2$ and CO. Furthermore, RCI exhibits robust performance and reaches chemical accuracy on the highly challenging iron-sulfur cluster using only 12\% of the full CI space. 
\hy{Notably, RCI outperforms recent regression-based SCI methods by delivering a more than 15\% improvement in accuracy at comparable determinant counts. RCI also demonstrates higher efficiency than heat-bath CI on the strongly correlated chromium dimer, yielding a compact and accurate wavefunction.} 
This pairwise learning-to-rank model provides a lightweight and modular plugin that can be seamlessly incorporated into other supervised learning frameworks.
\end{abstract}


\section{Introduction}

The accurate description of electron correlation remains a central challenge in computational chemistry, essential for predicting molecular properties with chemical accuracy. 
Full configuration interaction (FCI) expresses the wavefunction as a linear combination of the basis of all possible $N$-electron functions, namely Slater determinants (SDs).
While FCI provides the exact solution within a given one-particle basis set, its SD dimension grows factorially with molecular size, which limits its practical application.
To circumvent this bottleneck, selected configuration interaction (SCI) methods, such as CIPSI~\cite{CIPSI}, adaptive sampling CI~\cite{ASCI_1,ASCI_2}, heat-bath CI (HCI)~\cite{HCI_1,HCI_2,HCI_3} and others~\cite{directsci,PTCI,AMC,dCI}, have emerged as powerful alternatives. 
These methods approximate the FCI limit by iteratively selecting only the most variationally significant determinants from the vast Hilbert space. 
\hy{The practical viability of such truncated expansions is rooted in the smoothness properties of the many-electron wavefunction, which guarantee a rapid convergence of the CI expansion~\cite{griebel2007sparse,chinnamsetty2018adaptive,anderson2018breaking}.
Consequently, an accurate approximation to the FCI limit can be achieved with a remarkably compact subset of determinants, effectively circumventing the curse of dimensionality inherent to the full Hilbert space.}
Recently, the integration of machine learning (ML) into SCI has shown great promise in accelerating this selection process. 
By learning the mapping between determinant descriptors and their importance, ML models can bypass expensive perturbative estimates, acting as efficient prescreening tools to identify high-contribution configurations.

A prevalent strategy for identifying important configurations relies on supervised learning, where a model is trained to predict the importance of a determinant based on its wavefunction coefficient. This task is typically framed as either a \emph{regression} or a \emph{classification} problem.
In regression approaches~\cite{mlci,mlci_2,prob,cigen,shang,haar-sci}, a common practice is to first apply monotonic transformations to the CI coefficients---such as rescaling them to the $[0,1]$ interval or casting them as parameters of a probability distribution---and then fit these transformed targets via a regression loss.
Conversely, classification approaches~\cite{chembot,ALCI,bilous2023deep,bilous2024neural,bilous2025neural,nnci,nnci_NO,casier2026machine} label configurations as important or unimportant based on a predefined threshold, utilizing classification losses to align the predicted probabilities with these binary labels.
While existing regression and classification methods have accelerated CI selection, a fundamental limitation lies in the objective-loss mismatch. 
In the context of SCI, the primary objective is ranking: distinguishing which determinants are relatively more important than others to prioritize their inclusion in the variational space.
However, standard regression and classification objectives serve as intermediate proxy metrics that do not directly align with this ultimate importance metric of determinants.
Specifically, they rely on pointwise losses (e.g., mean squared error (MSE) or cross-entropy) in ranking that evaluate each determinant in isolation.
Regression methods tend to overemphasize minimizing the numerical deviation between predicted and true values rather than capturing the relative magnitude of coefficients in the subspace. This focus often leads to poor resolution for small but physically significant determinants, resulting in an inefficient expansion of the variational subspace. For instance, some unimportant determinants in the full expansion, which may not be negligible at all in the selected subspace, are not accounted for by pointwise regression.
On the other hand, classification approaches inherently discard the continuous nature of the wavefunction coefficients by imposing an artificial hard threshold. Consequently, the model fails to capture the fine-grained relative importance among the selected determinants, treating all important configurations as effectively equal.
Notably, \citeauthor{rlci}~\cite{rlci} introduced Reinforcement Learning CI, which employs a reinforcement learning strategy to maintain a weight vector spanning the full determinant space, effectively learning a global ranking. However, this ranking-first philosophy has not yet been systematically exploited within the supervised learning framework for SCI.

To bridge this gap, we draw inspiration from Information Retrieval~\cite{Liu2011} and propose a paradigm shift from learning coefficients to Learning to Rank (LTR).
Our approach, termed ranking configuration interaction (RCI), reframes the determinant selection process as a pairwise ranking problem, employing a Transformer-based architecture to capture complex non-local many-body dependencies through self-attention~\cite{transformer}.
Instead of optimizing a pointwise surrogate loss on individual data points, we employ a Pairwise Logistic Loss~\cite{ranknet}. 
While technically still a proxy metric, this pairwise formulation provides a significantly more rigorous alignment with the true ranking objective by explicitly optimizing the partial ordering.
This process is further augmented by an active pair sampling strategy with hard negative mining to accelerate model training.
By aligning the training objective with the actual selection criteria in SCI, our approach minimizes the mismatch between training and application, enabling the model to prioritize the most physically significant determinants more effectively than traditional regression or classification baselines. 
Extensive benchmark evaluations validate the effectiveness and advantages of this approach.
Across plane-wave basis sets, RCI reached \hy{correlation energies close to the reduced-space FCI reference} in \ce{N2} and \ce{CO} using only $\sim$55\% of the determinant count and less than 46\% of the wall time compared to the classification-based NNCI baseline~\cite{nnci}. 
Similar efficiency gains were observed with Gaussian basis sets (cc-pVDZ), where RCI reduced computational time by 23\%--40\% across \ce{N2}, \ce{C2}, \ce{H2O}, and \ce{NH3} while successfully converging to the exact FCI energies. 
Furthermore, RCI demonstrated robust performance on the strongly correlated \ce{N2} dissociation curve, achieving correlation energies $0.72$~mHa lower than NNCI on average while consuming only 71.5\% of the wall time. 
\hy{
RCI's efficiency advantage also persists as the orbital space grows: across three basis sets of increasing size on \ce{BeH2} (6-31G, cc-pVDZ, cc-pVTZ), it consistently achieved a more compact wavefunction than NNCI at similar accuracy, with no sign of diminishing returns even in the largest basis set.
}
On the highly challenging \ce{[Fe2S2(SCH3)4]^{2-}} cluster, RCI reached chemical accuracy (1.35~mHa error) using only $\sim$12\% of the full FCI space, outperforming the recent regression-driven Transformer SCI methods~\cite{shang,haar-sci} by more than 15\% in accuracy.
\hy{
On the substantially larger, strongly correlated \ce{Cr2} dimer, RCI further achieved both a lower variational and PT2-corrected energy than SHCI at matched determinant count.
}
Finally, ablation studies confirmed that the synergy between the Transformer architecture and the LTR objective is crucial for maximizing determinant selection efficiency, with the active pair sampling strategy further accelerating training by prioritizing maximally informative determinant pairs.

\section{Methods}

\subsection{SCI as a Ranking Problem}

\subsubsection{SCI Selection}

The nonrelativistic Hamiltonian of a many-electron system in the second quantization formalism reads:
\begin{equation}
    \hat{H} = \sum_{ij} h_{ij} a_i^{\dagger}a_j 
    + \frac{1}{2}\sum_{ijkl} \langle ij|kl \rangle \, a_i^{\dagger}a_j^{\dagger}a_l a_k
\end{equation}
where $a_i^{\dagger}$ and $a_i$ are Fermionic creation and annihilation operators with spin-orbital index $i$, and $h_{ij}$ and $\langle ij|kl \rangle$ are the one- and two-electron integrals, respectively.
The exact many-body ground state $\ket{\Psi_0}$ is defined as the eigenstate of the Hamiltonian $\hat{H}$ corresponding to its lowest eigenvalue $E_0$, satisfying
\begin{equation}
    \hat{H}\ket{\Psi_0} = E_0\ket{\Psi_0}
\end{equation}
This state can be expanded over orthonormal basis states of all allowed Slater determinants, $\{\phi_i\}$, that span the full Hilbert space, $\mathcal{H}$:
\begin{equation}
    \ket{\Psi_0} = \sum_{\ket{\phi_i} \in \mathcal{H}} c_i \ket{\phi_i}
\end{equation}
where $c_i$ denotes the CI coefficient associated with $\ket{\phi_i}$.
In practice, the dimension of $\mathcal{H}$ grows factorially with system size, rendering an exact FCI treatment computationally intractable for all but the smallest systems. SCI methods seek to approximate the FCI target state by iteratively constructing a variational subspace $\mathcal{V} \subset \mathcal{H}$ that captures the dominant contributions to the wavefunction. At each iteration, candidate determinants are generated according to $\hat{\mathcal{O}}\mathcal{V}$ out of the complementary space $\mathcal{H} \setminus \mathcal{V}$ by applying an extension operator $\hat{\mathcal{O}}$ to the determinants in the current variational space $\mathcal{V}$. 
A selection criterion is then applied to identify the most variationally significant candidates for inclusion in the next iteration.

Classical SCI methods such as CIPSI~\cite{CIPSI} and HCI~\cite{HCI_1,HCI_2,HCI_3} address this challenge by computing perturbative importance estimates to guide determinant selection. 
ML-supported SCI methods replace or augment this screening with a learned model $\mathcal{M}: |\phi_k\rangle \mapsto s_k$ that assigns a scalar importance score to the candidate determinant. 
Once trained, $\mathcal{M}$ can score all candidates in a single forward pass, bypassing the expensive Hamiltonian evaluations and perhaps diagonalization. 
The effectiveness of this approach therefore hinges critically on how well the model's training objective aligns with the true selection criterion, namely the relative ranking of selected determinants by their often varying significance as the subspace $\mathcal{V}$ grows. Finally, as the subspace $\mathcal{V}$ expands towards FCI, the contribution of individual determinants generally decreases as the wavefunction becomes highly complex, making importance scoring significantly less effective.

\subsubsection{LTR and Its Analogy to SCI}

LTR is a family of supervised machine learning techniques originally developed to train a ranking model that correctly sort items---either documents or search results---by their relevance for a specific query.
Formally, given a query $q$ and its feature vectors $\mathbf{x} = \{x_i\}_{i=1}^{m}$ representing $m$ candidate items labeled with ground-truth relevance $\mathbf{y}=\{y_i\}_{i=1}^{m}$, LTR learns a scoring function $f$ that outputs the most relevant order to the query.

Inspired by the LTR mechanism, the analogy between LTR and SCI is both direct and natural.
In our SCI setting,  the candidate determinants $\{\phi_k\}$ are sorted to match the ordering of their CI coefficients $\{|c_k|\}$ for the querying wavefunction $|\Psi_{\mathcal{V}}\rangle$. 
The $|\Psi_{\mathcal{V}}\rangle$, $\{\phi_k\}$ and $\{|c_k|\}$ correspond exactly to the LTR ``query'', ``items'' and ``relevance labels'', respectively,
\begin{equation}
    |\Psi_{\mathcal{V}}\rangle \mapsto q, \quad \{\phi_k\} \mapsto \mathbf{x}, \quad \{|c_k|\} \mapsto \mathbf{y}.
\end{equation}
The SCI selection criterion, admitting the top-$K$ most important determinants into $\mathcal{V}$, is precisely a top-$K$ retrieval task in which only the relative ordering of candidates matters, not their absolute scores.
This correspondence motivates our core contribution: a reconceptualization of ML-supported SCI as a ranking problem rather than a regression or classification task.

\hy{
LTR approaches are broadly categorized into three paradigms based on their input space, output space, and loss function~\cite{Liu2011}:}
\begin{itemize}
\item \hy{Pointwise methods treat each document $x_i$ independently as the input, defining the output as a scalar relevance score $y_i$~\cite{pointwise}. The loss $\mathcal{L}(f;\, x_i, y_i)$ penalizes inaccurate relevance prediction via regression, classification, or ordinal regression, and is applied to each document in isolation.}

\item \hy{Pairwise methods operate on document pairs $(x_i, x_j)$ as input, defining the output as the pairwise preference~\cite{ranknet,lambdarank}:
\begin{equation}
    y(i,j) = 2 \cdot \mathbb{I}[y_i > y_j] - 1 \in \{+1, -1\}.
\end{equation}
The loss $\mathcal{L}(f;\, x_i, x_j, y(i,j))$ penalizes inconsistencies between predicted and ground-truth preferences, typically expressed in terms of the score difference $f(x_i) - f(x_j)$.}

\item \hy{Listwise methods take the full document set $\mathbf{x} = \{x_i\}_{i=1}^{m}$ associated with a query as input, with the output defined as a ranked permutation $\pi_y$~\cite{listnet,listmle}. The loss $\mathcal{L}(f;\, \mathbf{x}, \pi_y)$ is defined over all documents jointly and cannot be decomposed into a sum over individual documents or pairs, making document positions explicitly visible to the objective.}
\end{itemize}

\hy{
To assess the effect of the loss function, we evaluated several representative candidates from the pairwise and listwise paradigms.
For pairwise losses, we tested RankNet's plain logistic loss~\cite{ranknet} and its LambdaRank variant~\cite{lambdarank}. The latter additionally weights each pair by the gain in Normalized Discounted Cumulative Gain (NDCG)~\cite{dcg} induced by swapping the two items, thereby placing greater emphasis on the top of the ranking.
For listwise losses, we tested ListMLE~\cite{listmle}, which maximizes the log-likelihood of the ground-truth ranking under the Plackett--Luce model defined by the predicted scores~\cite{plackett1975analysis}.
The results show that pairwise losses outperform ListMLE (see Supporting Information for details), which may be attributed to the large candidate set and highly skewed relevance distribution characteristic of the SCI setting.
Among the pairwise losses, LambdaRank achieves slightly lower energies in early iterations but exhibits a less stable convergence trajectory compared to RankNet.
Therefore, in this study, we adopt the plain logistic loss of RankNet and compare it against established classification and regression baselines.
}

\subsection{RCI Algorithm}

RCI builds upon the iterative subspace expansion framework recently proposed by \citeauthor{nnci}~\cite{nnci} in the context of neural-network configuration interaction (NNCI). 
The open-source SOLAX library~\cite{solax} serves as the computational backbone, providing a versatile infrastructure for handling basis sets, wavefunctions, and operators. 
For benchmarking on small molecules, such as \ce{N2} and \ce{CO}, RCI follows the NNCI approach by explicitly constructing and storing the Hamiltonian matrix, followed by direct diagonalization.
For large-scale systems, such as the \ce{[Fe2S2(SCH3)4]^{2-}} cluster, RCI instead employs Davidson diagonalization implemented via modified PySCF subroutines~\cite{sun2018pyscf, sun2020recent}.
All deep learning components of the architecture are implemented within the PyTorch framework~\cite{pytorch}.
The default hyperparameter settings of RCI are provided in the Supporting Information; unless otherwise stated, all experiments are conducted using these default settings.

The core workflow follows an active learning cycle designed to iteratively identify and include the most variationally significant SDs from the vast Hilbert space. As illustrated in Figure~\ref{fig:main}, the process proceeds through seven distinct stages in each iteration $t$:

\begin{figure*}[ht]
\centering 
\includegraphics[width=0.95\textwidth]{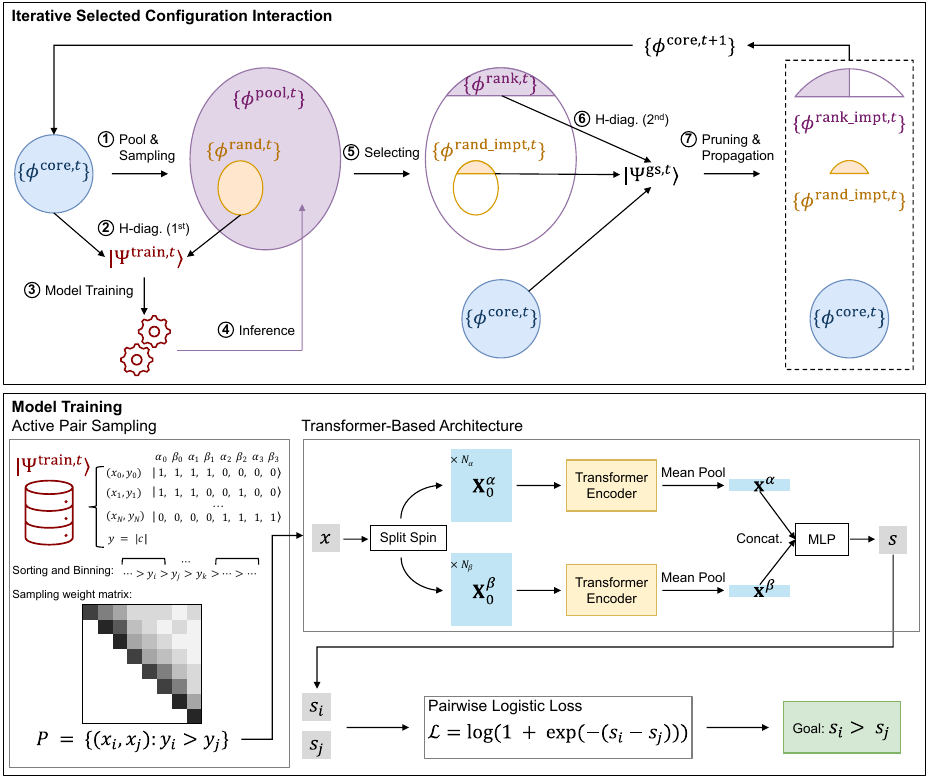} 
\caption{
Overview of the RCI framework. 
The upper panel illustrates the workflow of the iterative selected configuration interaction procedure. 
The lower panel details the third step, model training, where a Transformer-based model is trained with a pairwise Learning to Rank objective to predict determinant importance.
}
\label{fig:main}
\end{figure*}

\begin{enumerate}
    \item \textbf{Pool and Sampling:} Iteration $t$ commences with a core variational space, denoted as $\{\phi^{\text{core}, t}\}$, which comprises the most significant determinants identified in the preceding steps. 
    For the initial step ($t=0$), this core space is initialized by applying an extension operator $\hat{\mathcal{O}}$ twice to the Hartree--Fock reference determinant.
    Following NNCI, this extension operator truncates the full Hamiltonian to retain only dominant excitations (i.e., integral amplitudes $\ge 0.02$).
    Thus, the initial core is analogous to a selected CI singles and doubles (CISD).
    From this core space, a broader set of candidate determinants, the ``pool'', is generated via the extension operator: $\{\phi^{\text{pool}, t}\} = \hat{\mathcal{O}}\{\phi^{\text{core}, t}\} \setminus \{\phi^{\text{core}, t}\}$.
    Next, a random subset of the pool, denoted as $\{\phi^{\text{rand}, t}\}$, is sampled to construct the training space. 
    Adopting the same heuristic as in NNCI, the sampling ratio $p^{\text{rand},t}$ is set to be inversely proportional to the square root of the total pool size.
    
    \item \textbf{Hamiltonian Diagonalization (1st):} The Hamiltonian matrix is constructed within the subspace spanned by the union of $\{\phi^{\text{core}, t}\}$ and $\{\phi^{\text{rand}, t}\}$. Subsequent diagonalization yields the intermediate wavefunction $\ket{\Psi^{\text{train}, t}}$, whose CI coefficients reflect the relative importance of determinants, thereby serving as the supervision signal for model training.
    
    \item \textbf{Model Training:} A Transformer-based neural network is trained using the labeled state $\ket{\Psi^{\text{train}, t}}$. 
    Unlike the standard classification or regression strategies in previous studies, RCI formulates this step as an LTR problem. 
    Specifically, we employ a pairwise ranking objective, which exhibits robust and stable convergence behavior. 
    To further accelerate training, an active pair sampling strategy is introduced. 
    The sampling and training details of this step are provided in Section~\nameref{sec:pairwise_ltr}.

    \item \textbf{Inference:} The trained model is then deployed to scan the vast candidate pool ($\{\phi^{\text{pool}, t}\}$). It assigns a scalar importance score to each determinant, effectively predicting its potential contribution to the ground state.
    
    \item \textbf{Selecting:} Two distinct subsets are identified for inclusion in the variational space. 
    From the random training samples ($\{\phi^{\text{rand}, t}\}$), determinants with coefficient magnitudes exceeding a predefined threshold are extracted to form $\{\phi^{\text{rand\_impt}, t}\}$.
    From the remaining candidate pool ($\{\phi^{\text{pool}, t}\} \setminus \{\phi^{\text{rand}, t}\}$), the model's predicted importance scores are utilized to select the highest-ranking determinants, denoted as $\{\phi^{\text{rank}, t}\}$. The specific selection protocol, including dynamic thresholding, is detailed in Section~\nameref{sec:selection_strat}.
    
    \item \textbf{Hamiltonian Diagonalization (2nd):} A second, comprehensive diagonalization is performed on the augmented subspace, which now encompasses the original core ($\{\phi^{\text{core}, t}\}$), the determinants selected by the model ($\{\phi^{\text{rank}, t}\}$), and the significant random samples ($\{\phi^{\text{rand\_impt}, t}\}$). This step yields the ground state wavefunction $\ket{\Psi^{\text{gs}, t}}$ and the corresponding energy estimate for the current iteration.
    
    \item \textbf{Pruning and Propagation:} To maintain computational efficiency for subsequent iterations, the variational space undergoes a pruning process. 
    Specifically, determinants in $\{\phi^{\text{rank}, t}\}$ whose coefficients in $\ket{\Psi^{\text{gs}, t}}$ fall below the threshold defined in Step~5 are discarded. 
    This step eliminates false positive determinants that proved insignificant upon the second diagonalization in Step~6. 
    The remaining significant determinants are aggregated to form the updated core space, $\{\phi^{\text{core}, t+1}\}$, which is then propagated to the next iteration.

\end{enumerate}

\begin{figure}[ht]
\centering 
\includegraphics[width=0.8\textwidth]{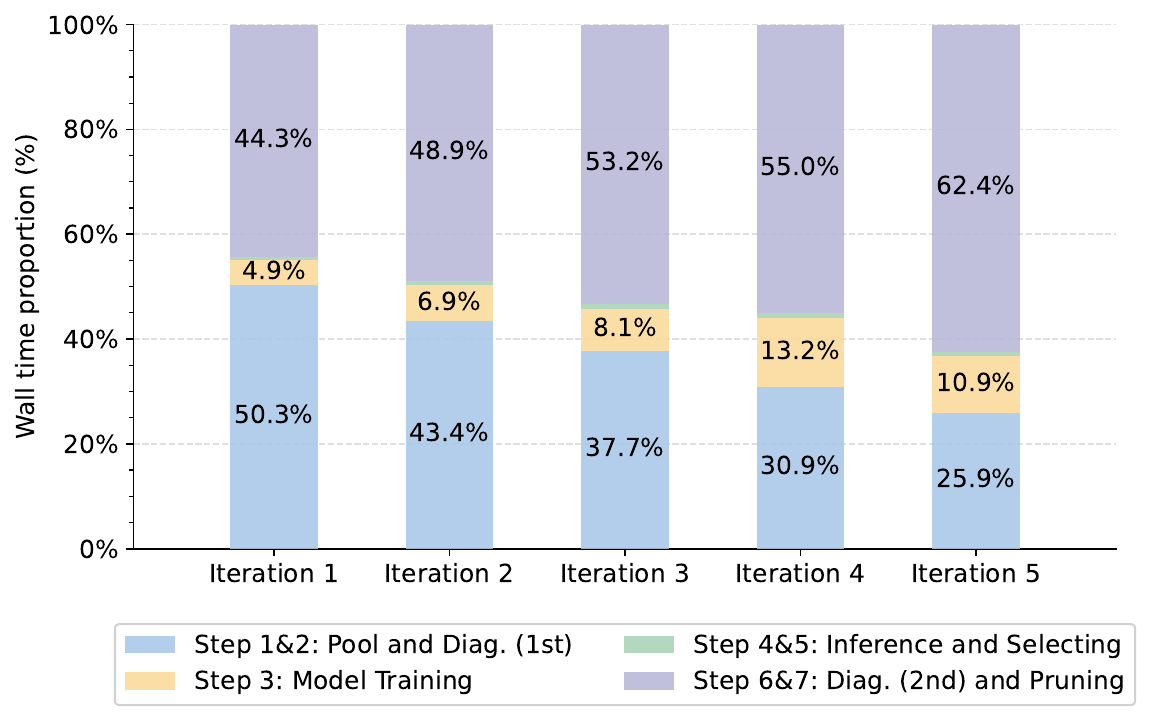} 
\caption{
Wall time proportion of each computational stage across RCI iterations for \ce{N2}. 
Each bar represents one iteration, decomposed into four stages: Steps~1\&2 (pool construction and first Hamiltonian diagonalization), Step~3 (model training), Steps~4\&5 (model inference and determinant selection), and Steps~6\&7 (second diagonalization and pruning). 
Steps~4\&5 account for less than 1\% of the total wall time and are not labeled with numerical annotations.
}
\label{fig:time_frac}
\end{figure}

Timing analysis on the plane-wave basis \ce{N2} example (CAS(10e,52o)) demonstrates that the computational cost is dominated by Hamiltonian diagonalization, which accounts for the majority of wall time in each iteration (Figure~\ref{fig:time_frac}). 
In contrast, model training with the active pair sampling strategy and model inference remain lightweight throughout the iterative procedure, never exceeding 15\% and 1\% of the total wall time per iteration, respectively. 
This indicates that despite adopting a more expressive Transformer architecture and a pairwise ranking objective, the additional training overhead introduced by RCI remains acceptable and does not constitute a computational bottleneck in the overall SCI workflow.

\subsection{Model Training}
\label{sec:pairwise_ltr}

The key difference between our approach and standard NNCI lies in both the model architecture and training objective. 
To better capture the characteristics of determinants, we employ a Transformer-based architecture that leverages self-attention to model complex inter-orbital dependencies.
On the other hand, to overcome the objective-loss mismatch inherent in conventional regression or classification approaches, we introduce the LTR paradigm. 
By employing a pairwise loss to optimize the score gap between determinant pairs, this approach provides a tighter proxy for the relative ordering than fitting CI coefficient values or binary importance labels. 
The complete training procedure is summarized in Algorithm~\ref{alg:train_pairwise}.

\begin{algorithm}[H]
\caption{Training procedure for the $t$-th iteration}
\label{alg:train_pairwise}
\begin{algorithmic}[1]
\Require Model $\mathcal{M}$ with parameters $\theta$,
train state $\ket{\Psi^{\text{train},t}}$,
batch size $N_{\text{pair}}$,
probe size $N_{\text{probe}}$,
number of bins $N_{\text{bins}}$,
tolerance $\epsilon$,
max epochs $N_\text{epoch}$,
update interval $N_\text{update}$,
evaluation interval $N_\text{eval}$,
learning rate $\eta$,
patience $N_\text{estop}$.

\State \textbf{Data preparation:}
\State $X_{\text{train}} \leftarrow \{\phi^{\text{core},t}\} \cup \{\phi^{\text{rand},t}\}$, \quad $Y_{\text{train}} \leftarrow \{|c_i| : \phi_i \in X_{\text{train}}\}$
\State $X_{\text{val}} \leftarrow \{\phi^{\text{rand},t}\}$, \quad\quad\quad\quad $Y_{\text{val}} \leftarrow \{|c_i| : \phi_i \in X_{\text{val}}\}$

\State $\mathcal{B} \leftarrow \text{Partition}(X_{\text{train}}, Y_{\text{train}}, N_{\text{bins}})$ 

\State $W \leftarrow \text{Uniform}(N_{\text{bins}} \times N_{\text{bins}})$ \Comment{Initialize bin pair weights}
\State $\rho_{\text{best}} \leftarrow -1.0,\; \mathcal{M}_{\text{best}} \leftarrow \mathcal{M},\; \text{stall} \leftarrow 0$

\State \textbf{Training loop:}
\For{$e = 1$ \textbf{to} $N_\text{epoch}$}
    \If{$e \bmod N_\text{update} = 0$}
        \State $W \leftarrow \text{UpdateWeight}(\mathcal{B}, \mathcal{M}, N_\text{probe})$ 
    \EndIf
    
    \State $P \leftarrow \text{SamplePair}(\mathcal{B}, W, N_{\text{pair}})$ 
    
    \State $\mathcal{L} \leftarrow \frac{1}{|P|} \sum_{(x_i,x_j) \in P} \log\left( 1 + \exp\left(-(\mathcal{M}(x_i) - \mathcal{M}(x_j))\right) \right)$
    
    \State $\theta \leftarrow \theta - \eta \nabla_{\theta} \mathcal{L} $

    \If{$e \bmod N_\text{eval} = 0$} \Comment{Validation}
        \State $S_{\text{val}} \leftarrow \mathcal{M}(X_{\text{val}})$  
        \State $\rho \leftarrow \text{Spearman}(S_{\text{val}}, Y_{\text{val}})$ 
        \If{$\rho > \rho_{\text{best}} + \epsilon$}
            \State $\rho_{\text{best}} \leftarrow \rho,\; \mathcal{M}_{\text{best}} \leftarrow \mathcal{M},\; \text{stall} \leftarrow 0$
        \Else
            \State $\text{stall} \leftarrow \text{stall} + 1$
        \EndIf
        \If{$\text{stall} \ge N_\text{estop}$}
            \State \textbf{break}
        \EndIf
    \EndIf
\EndFor
\State \Return $\mathcal{M}_\text{best}$
\end{algorithmic}
\end{algorithm}

\subsubsection{Transformer-Based Architecture}

To effectively capture the complex correlations between electronic configurations, we employ a dual-path Transformer-based~\cite{transformer} neural network architecture. 

\begin{itemize}
\item \textbf{Input Representation:}
The input to the model is a determinant represented as an interleaved bitstring $x = (\alpha_1, \beta_1, \ldots, \alpha_{N_{\text{orb}}}, \beta_{N_{\text{orb}}})$, where $\alpha_i, \beta_i \in \{0, 1\}$ indicate the occupation status of the $i$-th spatial orbital for alpha and beta spins, respectively, and $N_{\text{orb}}$ denotes the total number of spatial orbitals.
Let $y_k$ denote the magnitude of the CI coefficient associated with a determinant $x_k$, i.e., $y_k = |c_k|$.

\item \textbf{Embedding Layer:}
To vectorize this discrete representation, we maintain two learnable embedding matrices for the alpha and beta spin channels: $\mathbf{E}^\alpha \in \mathbb{R}^{N_{\text{orb}} \times d}$ and $\mathbf{E}^\beta \in \mathbb{R}^{N_{\text{orb}} \times d}$, where $d$ is the embedding dimension.
Specifically, we first decompose $x$ into separate spin components by extracting the indices of occupied orbitals. 
Let $\mathcal{I}^\alpha_x = \{i \mid \alpha_i = 1\}$ and $\mathcal{I}^\beta_x = \{i \mid \beta_i = 1\}$ be the strings of occupied orbitals for alpha and beta electrons, with cardinalities $|\mathcal{I}^\alpha_x| = N_\alpha$ and $|\mathcal{I}^\beta_x| = N_\beta$, where $N_\alpha$ and $N_\beta$ are the number of alpha and beta electrons.
Using these indices to look up the corresponding rows in $\mathbf{E}^\alpha$ and $\mathbf{E}^\beta$, we obtain two dense feature matrices:
\hy{
\begin{equation}
    \mathbf{X}^\alpha_0 \in \mathbb{R}^{N_\alpha \times d} \quad \text{and} \quad \mathbf{X}^\beta_0 \in \mathbb{R}^{N_\beta \times d}
\end{equation}
}
which encode the spatial orbital information for the alpha and beta electrons of determinant $x$, respectively.

\item \textbf{Encoder and Scoring:}
These feature matrices are processed by two independent Transformer encoders to capture non-local orbital interactions within each spin channel via self-attention mechanisms.
Following the encoding stage, a mean pooling operation is applied across the sequence dimension (i.e., over the $N_\alpha$ or $N_\beta$ electrons), resulting in two fixed-size feature vectors \hy{$\mathbf{x}^\alpha, \mathbf{x}^\beta \in \mathbb{R}^{d}$}.
These vectors are concatenated to form a unified state representation \hy{$[\mathbf{x}^\alpha; \mathbf{x}^\beta] \in \mathbb{R}^{2d}$}.
Finally, a Multi-Layer Perceptron (MLP) projection head maps \hy{$[\mathbf{x}^\alpha; \mathbf{x}^\beta]$} to a single scalar output $s$, representing the predicted importance score used for ranking.

\end{itemize}

\hy{
We further evaluated a fused-spin-channel variant that permits cross-spin attention.
Rather than processing the alpha and beta strings through two independent encoders, the occupied orbital embeddings from both spin channels are concatenated into a single sequence of length $N_\alpha + N_\beta$ before being passed to a shared Transformer encoder.
To preserve spin identity, a learnable spin-channel embedding $\mathbf{E}^{\text{channel}} \in \mathbb{R}^{2 \times d}$ is added to each token prior to encoding, analogous to segment embeddings in BERT~\cite{bert}.
The encoder output is then mean-pooled over the full fused sequence and mapped to a scalar score via the MLP projection head.
This design permits self-attention across alpha and beta electrons, explicitly modeling opposite-spin correlations.
}

\hy{
The fused-spin-channel variant achieves a convergence trajectory comparable to the dual-channel baseline (see Supporting Information for details).
Given that fusing the two spin channels into a single sequence of length $N_\alpha + N_\beta$ increases the self-attention cost from $\mathcal{O}(N_\alpha^2 + N_\beta^2)$ to $\mathcal{O}((N_\alpha + N_\beta)^2)$ and incurs higher memory consumption, we retain the dual-channel architecture as the default for RCI.
Nevertheless, this fused-spin-channel variant remains a viable and interchangeable alternative within our framework, which may be preferable for users with sufficient computational resources or for systems where explicit opposite-spin attention is architecturally desirable.
}

\subsubsection{Pairwise Logistic Loss}

We construct a set of training determinant pairs $P = \{(x_i, x_j) : y_i > y_j\}$ from the training state $\ket{\Psi^{\text{train}, t}}$, where each pair consists of two determinants such that the coefficient magnitude of $x_i$ strictly exceeds that of $x_j$. The objective is to train the model to assign a higher predicted score to the more important determinant in each pair, thereby learning the correct relative ordering.

We employ the Pairwise Logistic Loss, which directly optimizes the ranking objective by maximizing the probability that $x_i$ is ranked above $x_j$. Formally, the loss is defined as:
\begin{equation}
    \label{eq:loss}
    \mathcal{L} = \frac{1}{|P|} \sum_{(x_i,x_j) \in P} \log\left( 1 + e^{-(s_i - s_j)} \right)
\end{equation}
where $s_i$ and $s_j$ are the scalar scores predicted by the model for determinants $x_i$ and $x_j$, respectively. 
This loss function penalizes violations of the desired ranking order: when the model incorrectly assigns a higher score to the less important determinant ($s_j > s_i$), the exponential term $e^{-(s_i - s_j)}$ becomes large, resulting in a higher loss. Conversely, when the model correctly predicts $s_i > s_j$, the loss approaches zero. This formulation naturally enforces monotonicity in the learned ranking and is differentiable, enabling efficient gradient-based optimization.

\subsubsection{Active Pair Sampling}

Motivated by the observation that adaptive sampling distributions consistently outperform fixed strategies in metric learning~\cite{roth2020pads}, we implement an active sampling strategy based on coefficient magnitudes to improve sample efficiency and focus training on informative determinant pairs.
The determinants in the training state ($\ket{\Psi^{\text{train},t}}$) are first sorted in descending order by their coefficient magnitudes and partitioned into $N_\text{bins}$ bins. 
The sampler constructs pairs from two categories: (1) distant bin pairs, formed between determinants from distantly separated bins, which are easier to distinguish due to their larger coefficient differences; and (2) proximal bin pairs, formed within the same or nearby bins, which are more challenging as their coefficient magnitudes are nearly identical and thus require finer-grained discrimination.

We maintain a sampling weight matrix $W \in \mathbb{R}^{N_{\text{bins}} \times N_{\text{bins}}}$ to govern the selection probability of bin pairs. 
Since the determinants are sorted, a valid pair $(x_i, x_j)$ satisfying $y_i > y_j$ implies that the bin index of $x_i$ must be less than or equal to that of $x_j$. 
Consequently, $W$ is structured as an upper triangular matrix over the set of valid bin pairs $\triangle = \{(a,b) : a \leq b\}$, where entries near the diagonal correspond to proximal bin pairs, while far off-diagonal entries correspond to distant bin pairs.
Initially, the non-zero entries of $W$ are initialized uniformly. 
To facilitate hard negative mining, these weights are updated dynamically every $N_\text{update}$ training steps.

Specifically, we sample $N_\text{probe}$ determinant pairs $P_{ab}$ from each bin pair $(a, b) \in \triangle$. 
The average pairwise ranking loss for each bin pair is then computed as $L_{ab} = \mathcal{L}(P_{ab})$, where the loss $\mathcal{L}$ is defined in Eq.~\ref{eq:loss}. 
To amplify the contrast between easy and hard bin pairs, the raw losses are raised to a temperature exponent $\tau > 1$: $\tilde{L}_{ab} = \left(L_{ab}\right)^{\tau}$. 
These rescaled losses are then normalized into a sampling probability distribution; a small uniform mixture weighted by $\varepsilon$ is added to ensure every bin pair retains a minimum probability of being sampled:
\begin{equation}
    p_{ab} = (1 - \varepsilon)\cdot\frac{\tilde{L}_{ab}}{\displaystyle\sum_{(a',b')\in\triangle} \tilde{L}_{a'b'}} + \frac{\varepsilon}{|\triangle|}.
\end{equation}
Finally, the weight matrix is updated via an exponential moving average with decay $\lambda$:
\begin{equation}
    W_{ab} \leftarrow (1 - \lambda)\, W_{ab} + \lambda\, p_{ab}.
\end{equation}
This adaptive scheme ensures that the model progressively concentrates training on the most difficult ranking boundaries.

\subsubsection{Early Stopping Strategy}

We evaluate the model's ranking performance using the Spearman rank-order correlation coefficient, $\rho$, between the predicted importance scores $s$ and the ground-truth coefficient magnitudes $y$.
To ensure robust generalization to the unseen Hilbert space, we implement a distribution-aware early stopping strategy.
Although the model is trained on the full set of determinants of $\ket{\Psi^{\text{train},t}}$, namely $\{\phi^{\text{core},t}\} \cup \{\phi^{\text{rand},t}\}$, we monitor $\rho$ exclusively on the $\{\phi^{\text{rand},t}\}$ subset for model selection.
This approach is motivated by the distinct roles of the two subsets: $\{\phi^{\text{core},t}\}$ provides strong, high-magnitude signals that stabilize training, whereas $\{\phi^{\text{rand},t}\}$ is representative of the vast, unseen pool, characterized by a dominance of sparse, low-magnitude determinants.
Standard validation on the full set would be heavily biased by the high-signal $\{\phi^{\text{core},t}\}$. By restricting validation to $\{\phi^{\text{rand},t}\}$, we ensure the model retains the ability to generalize to these sparse, unexplored regions, rather than being overfitted to the strong signals of the known core determinants.

\subsection{Dynamic Selection Strategy}
\label{sec:selection_strat}

To identify the most variationally significant determinants for the augmented subspace, we employ a dual-selection strategy. This process treats the training subset and the unexplored pool differently, leveraging the specific information available in each case:

\begin{enumerate}
    \item \textbf{Selection from Random Samples ($\{\phi^{\text{rand}, t}\}$):} Leveraging the CI coefficients obtained from the preliminary Hamiltonian diagonalization, we establish a coefficient threshold corresponding to the top $p^{\text{impt},t}$ fraction of determinants in $\{\phi^{\text{rand}, t}\}$. Determinants exceeding this threshold are subsequently selected to constitute the subset $\{\phi^{\text{rand\_impt}, t}\}$. We adopt the same iteration-dependent schedule for $p^{\text{impt},t}$ as established in the standard NNCI protocol.
    
    \item \textbf{Selection from the Remaining Pool ($\{\phi^{\text{pool}, t}\} \setminus \{\phi^{\text{rand}, t}\}$):} For the vast majority of candidates where coefficients are unavailable, we rely on the trained model's predicted ranking scores. 
    The top $p^{\text{rank},t}$ fraction of candidates ordered by predicted score are selected to form the subset $\{\phi^{\text{rank}, t}\}$.
\end{enumerate}

Crucially, we observed that the model's ranking fidelity tends to diminish in later iterations as the variational space becomes more complex and the distinction between determinants becomes more subtle. 
To compensate for this increasing uncertainty in the model's predictions, we implement a dynamic inflation of the selection ratio specifically for the remaining pool:
\hy{
\begin{equation}\label{eq:select}
    p^{\text{rank},t} = \max\!\Big(p^{\text{impt},t},\ \min\!\big(p^{\text{cap}},\ p^{\text{impt},t} \cdot \gamma^{t-1}\big)\Big)
\end{equation}
The inflation factor $\gamma \geq 1$ progressively widens the selection window relative to the importance-sampling baseline $p^{\text{impt},t}$, while $p^{\text{cap}}$ prevents runaway growth. 
The outer $\max$ guards against the edge case where $p^{\text{impt},t}$ already exceeds $p^{\text{cap}}$, ensuring the cap only becomes active once the inflation factor $\gamma^{t-1}$ has grown large enough.
By gradually relaxing the selection threshold over iterations, potentially significant determinants are less likely to be discarded due to minor ranking errors in deeper, more entangled regions of the variational space.
}

\hy{
As illustrated in the Supporting Information, the pool size grows approximately exponentially as the iterations proceed, while $p^{\text{impt},t}$, which scales inversely with the square root of the pool size, decreases correspondingly.
Owing to the inflation factor $\gamma$, $p^{\text{rank},t}$ adaptively increases until it reaches the cap ratio $p^{\text{cap}}$ in later iterations.
The parameter sensitivity analysis reveals a clear division of roles between the two hyperparameters: 
the inflation factor $\gamma$ primarily governs convergence behavior in the early iterations before $p^{\text{cap}}$ is reached, where a smaller value enforces a more compact wavefunction by slowing the expansion of the variational space; 
the cap $p^{\text{cap}}$, by contrast, takes effect only once the inflation term grows large enough to exceed it, after which a smaller cap imposes greater compactness in the later iterations.
}

\section{Results}

In this section, we systematically evaluate the effectiveness and robustness of the proposed RCI approach. 
All benchmark experiments were conducted on a high-performance computing node equipped with dual-socket Intel Xeon Gold 6330 CPUs (56 physical cores, 2.00~GHz) and two NVIDIA A800 80~GB PCIe GPUs. 
The experimental results are organized as follows:

First, we assess the accuracy of recovering correlation energies at equilibrium geometries across both plane-wave and Gaussian basis sets. 
In this evaluation, we explicitly compare RCI with NNCI~\cite{nnci}---a recent ML-supported SCI method based on classification---to demonstrate the advantages of our LTR strategy over traditional classification approaches. 
We acknowledge the open-source implementation of NNCI, which facilitated a direct and reproducible comparison.
Second, to assess the robustness of the model when applied to strongly correlated, multi-reference systems beyond equilibrium geometries, we examine the potential energy curve (PEC) of the nitrogen molecule (\ce{N2}) during bond dissociation.
Third, we further extend our approach by applying RCI to the highly challenging \ce{[Fe2S2(SCH3)4]^{2-}} transition metal cluster.
Here, we benchmark our results against high-fidelity DMRG calculations to demonstrate the scalability and accuracy of RCI within massive Hilbert spaces.
Finally, we conduct comprehensive ablation studies to isolate and analyze the impact of specific algorithmic components, particularly the synergy between the Transformer architecture and the LTR objective, as well as the efficacy of our active pair sampling strategy.

\subsection{Correlation Energy at Equilibrium Geometries}

Here, we benchmark the efficiency of RCI and NNCI in recovering ground-state correlation energies for various small molecules at equilibrium, spanning both plane-wave and Gaussian basis sets.
Specifically, the number of selected determinants and the total computational wall time required to reach \hy{FCI reference energies} are systematically compared.
\hy{For consistency, we report the variational energy of the full wavefunction $\ket{\Psi^{\text{gs}, t}}$ before pruning for both RCI and NNCI throughout this comparison.}
We adopt this metric because it represents the true variational upper bound achieved at each iteration, fully reflecting the contribution of all selected determinants for which the computational cost of diagonalization has already been incurred.

\subsubsection{Results on Plane-Wave Basis Sets}
We first evaluate the performance of RCI on four representative molecules: \ce{N2}, \ce{CO}, \ce{H2O}, and \ce{NH3}. 
One- and two-electron integrals were collected from the NNCI study~\cite{nnci}, which utilized a plane-wave basis with the projector augmented-wave (PAW) method~\cite{paw1,paw2}.
The molecular geometry for \ce{N2} was taken from Ref.~\citenum{fci_N2}, while the geometries for the other molecules were adopted from Ref.~\citenum{fci_others}.
The number of molecular orbitals (MOs) included in these systems is 52 for \ce{N2}, 46 for \ce{CO}, 43 for \ce{H2O}, and 49 for \ce{NH3}. 

\begin{figure*}[ht]
\centering 
\includegraphics[width=\textwidth]{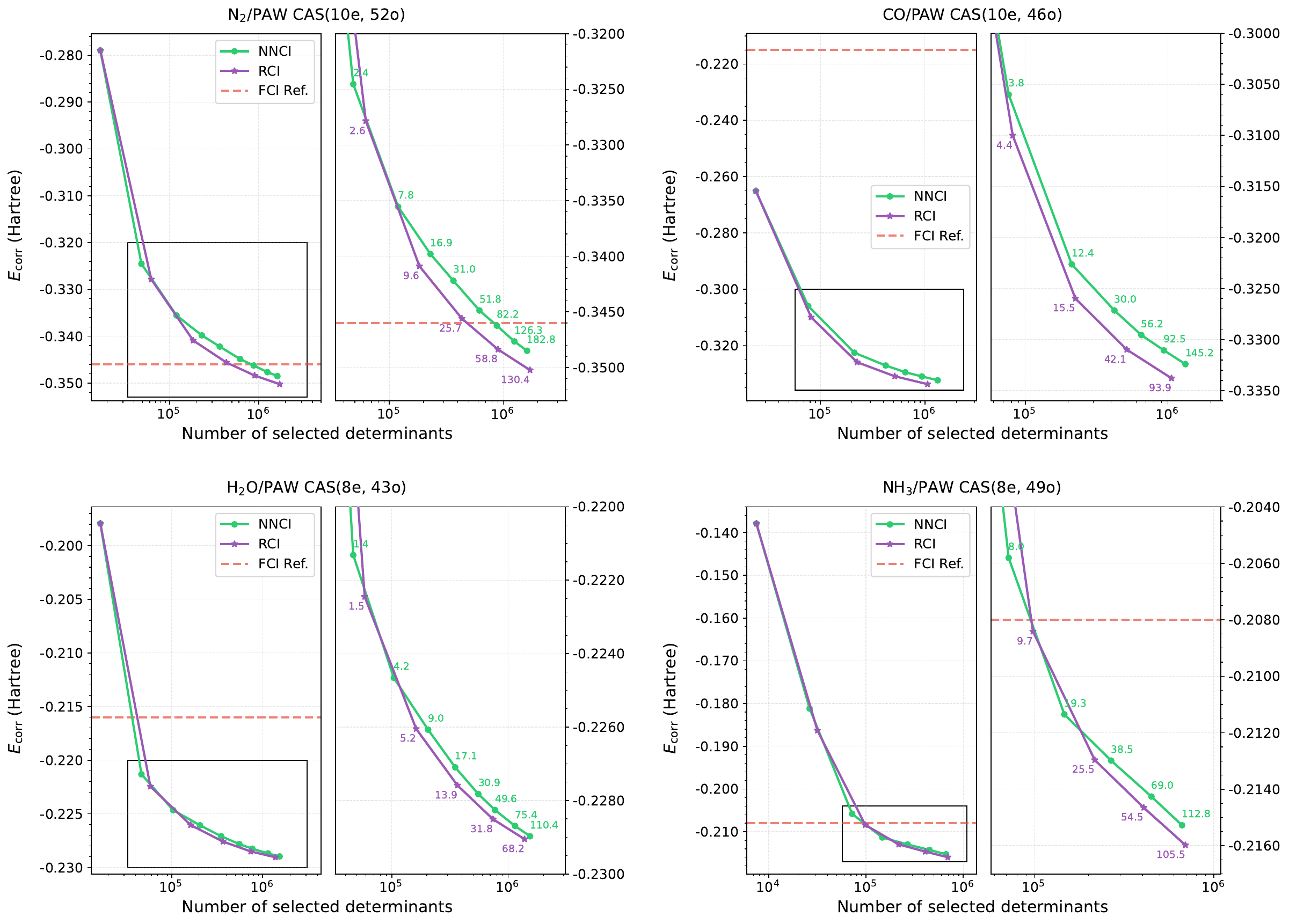} 
\caption{
Comparison of RCI and NNCI on molecular systems with plane-wave basis.
The recovered correlation energy is shown as a function of the number of selected determinants in the variational space for \ce{N2}, \ce{CO}, \ce{H2O}, and \ce{NH3}. 
Lower curves indicate more efficient recovery of correlation energy with fewer determinants. 
Selected wall times are annotated near the corresponding data points, in hours. 
The reference FCI energies (FCI Ref.) for \ce{N2}~\cite{fci_N2}, \ce{CO}~\cite{fci_others}, \ce{H2O}~\cite{fci_others}, and \ce{NH3}~\cite{fci_others} are taken from previously reported calculations with a smaller number of MOs.
}
\label{fig:plane_wave}
\end{figure*}

Figure~\ref{fig:plane_wave} illustrates the recovered correlation energy as a function of the variational space size (number of determinants). Due to the large number of MOs in these systems, exact FCI calculations are computationally prohibitive. Consequently, we benchmark our results against published FCI reference values derived from a reduced number of MOs~\cite{fci_N2,fci_others}, following the protocol established in the NNCI study.

As shown in Figure~\ref{fig:plane_wave}, the RCI convergence curves consistently lie below those of NNCI, indicating that RCI recovers a comparable amount of correlation energy using significantly fewer determinants. 
Taking \ce{N2} as an example, to achieve a correlation energy of approximately $-0.3485$ Ha, RCI requires only 55.6\% of the determinant count and 32.2\% of the wall time compared to NNCI. At the final iteration step, the RCI energy is $1.73$ mHa lower than that of NNCI, while consuming only 71.3\% of the total computational time. 
Similar efficiency gains are observed for \ce{CO}. To reach a correlation energy of $-0.3310$ Ha, RCI uses 54.9\% of the determinants and 45.5\% of the time required by NNCI. At the last iteration step, RCI achieves an energy $1.38$ mHa lower than NNCI with a total time cost of 64.7\%.
For \ce{H2O}, at the final point where both methods reach approximately $-0.2290$ Ha, RCI reduces the total computational time to 61.8\% of that used by NNCI. 
Finally, in the case of \ce{NH3}, the final RCI energy is $0.70$ mHa lower than NNCI, with a slight time advantage (93.6\% of the NNCI time).

It is worth noting that RCI selects a larger batch of determinants per iteration than the standard NNCI protocol, which may raise the concern that its faster convergence might be simply a consequence of the larger per-step batch size. 
The results in Section~\nameref{sec:ablation} address this by evaluating all variants under the same dynamic selection strategy, thereby controlling differences in selection policy. 
The results show that despite adopting a larger batch size, the modified NNCI variant achieves a comparable final energy to the original NNCI only at the cost of $2.1\times$ as many determinants and greater wall time ($196.2$ vs.\ $182.8$ hours). 
This confirms that selecting more determinants per step is insufficient on its own; the efficiency of RCI stems from its ability to prioritize important determinants by pairwise ranking, enabling a more efficient expansion of the variational space.

\subsubsection{Results on Gaussian Basis Sets}
Next, we evaluate the effectiveness of RCI using Gaussian basis sets which are more localized at atoms than plane-wave basis functions. 
Four benchmark systems were considered: the nitrogen dimer (\ce{N2}, CAS(10e,26o)), the carbon dimer (\ce{C2}, CAS(8e,26o)), the water molecule (\ce{H2O}, CAS(8e,23o)), and ammonia (\ce{NH3}, CAS(8e,28o)), all using the cc-pVDZ basis set.
One- and two-body integrals were obtained from restricted Hartree--Fock (RHF) calculations using the PySCF package~\cite{sun2018pyscf,sun2020recent}.
A distinct advantage of these systems is the accessibility of exact FCI solutions. 
Unlike the plane-wave cases where the large orbital space precludes exact diagonalization, these Gaussian-based benchmarks allow us to directly verify convergence toward the true FCI limit.

\begin{figure*}[ht]
\centering 
\includegraphics[width=\textwidth]{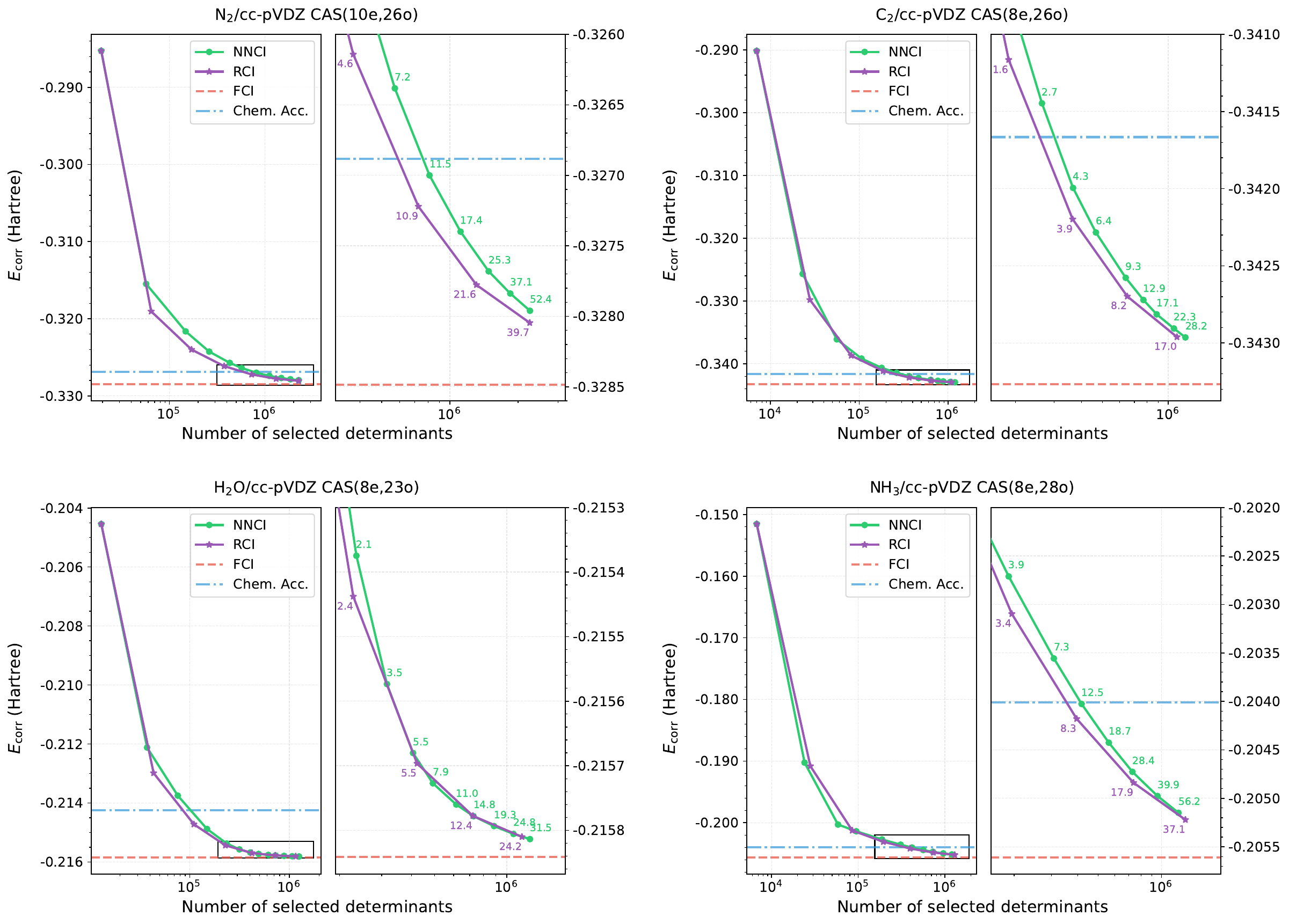} 
\caption{
Comparison of RCI and NNCI on Gaussian-basis benchmark systems: \ce{N2}, \ce{C2}, \ce{H2O}, and \ce{NH3} in the cc-pVDZ basis. 
The recovered correlation energy is plotted as a function of the number of selected determinants. 
The dashed horizontal lines in each panel denote the exact FCI correlation energy and the chemical accuracy (Chem.\ Acc.) threshold of 1.6 mHa. 
Selected wall times are annotated near the corresponding data points, in hours. 
}
\label{fig:gaussian}
\end{figure*}

As illustrated in Figure~\ref{fig:gaussian}, both RCI and NNCI successfully converge toward the exact FCI energy. 
Notably, the RCI convergence curve consistently lies below that of NNCI, demonstrating that RCI achieves comparable accuracy using a more compact wavefunction.
While both methods ultimately attain similar final correlation energies, RCI demonstrates superior computational efficiency. 
For \ce{N2} and \ce{NH3}, both methods converge to within $0.50$~mHa of the FCI limit, yet RCI consumes only 75.8\% and 66.6\% of the NNCI wall time, respectively. 
For \ce{C2} and \ce{H2O}, where the final errors are below $0.35$~mHa and $0.05$~mHa, respectively, RCI reduces the computational time to 60.3\% and 76.9\% of the NNCI cost.

In summary, results from both plane-wave and Gaussian basis sets consistently demonstrate the superior computational efficiency of RCI. This advantage stems from the effectiveness of the ranking mechanism, which accurately identifies the most significant determinants. By prioritizing high-value candidates, RCI constructs a compact yet accurate variational space, allowing for faster convergence to the \hy{exact FCI energies} compared to the classification-based NNCI approach.

\subsection{Dissociation Curve of \ce{N2}}

In this section, we evaluate the performance of RCI on the full dissociation curve of \ce{N2}, spanning bond lengths from $0.7$~\AA{} to $3.0$~\AA. 
The one- and two-electron integrals were sourced from the plane-wave/PAW calculations provided in the NNCI study~\cite{nnci}.
We benchmark our results against NNCI calculations using both 52 and 46 MOs.
Consistent with the NNCI methodology, we also compare our results to an experimental reference curve derived from a modified Morse + Lennard-Jones potential~\cite{N2_pec}, as well as to restricted Hartree--Fock (RHF) and FCI curves computed with smaller active spaces (18 and 28 MOs). 
All reference data were adopted directly from the NNCI publication.

\begin{figure*}[ht]
\centering 
\includegraphics[width=0.6\textwidth]{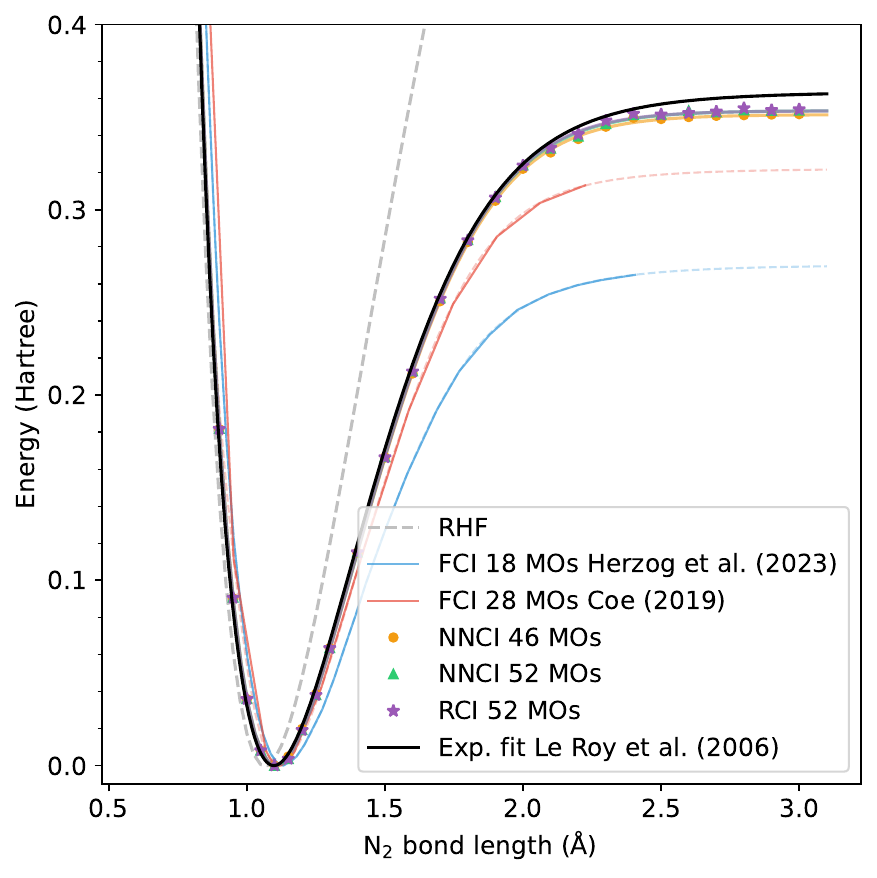} 
\caption{
Potential energy curves of \ce{N2} along the bond dissociation coordinate. 
RCI is compared with NNCI~\cite{nnci} calculations using different active orbital spaces, as well as with the same reference curves considered in the NNCI study, including RHF, reduced-space FCI results from \citeauthor{cigen}~\cite{cigen} (18 MOs) and \citeauthor{mlci_2}~\cite{mlci_2} (28 MOs), and the experimental (Exp.) fit reported by \citeauthor{N2_pec}~\cite{N2_pec} 
All curves are aligned at their minimum energy.
For consistency with NNCI, the RCI energies shown here correspond to the pruned wavefunction. 
The RCI curve exhibits improved agreement with the experimental reference, particularly in the stretched-bond regime ($> 2.1$~\AA) where strong correlation becomes more pronounced. 
}
\label{fig:dissociation}
\end{figure*}

The calculated potential energy curves are presented in Figure~\ref{fig:dissociation}. 
Both RCI and NNCI results significantly outperform the RHF and FCI (18/28 MOs) benchmarks, showing excellent agreement with the experimental curve from the equilibrium region to the dissociation limit.
Within the NNCI calculations, the use of 52 MOs yields a notably better curve than the 46-MO active space.
Moreover, RCI achieves correlation energies that are, on average, $0.72$~mHa lower than those of the best NNCI result (52 MOs), showing slightly better agreement with the experimental fit in the strongly correlated region (bond lengths $> 2.1$~\AA).
Crucially, this improved accuracy is achieved with significantly reduced computational effort: whereas NNCI required an average of 7 iterations per point, RCI converged in only 5 iterations.
Based on the benchmark at the equilibrium bond length ($1.111$~\AA; see Figure~\ref{fig:plane_wave}), this reduction in iteration count suggests that the total wall time of RCI is approximately 71.5\% of that required for NNCI (58.8 vs.\ 82.2 hours).

Regarding the equilibrium geometry, the experimental bond length ($a_{\text{min}}$) derived from Ref.~\citenum{N2_pec} is $1.0977$~\AA. 
Despite the improvement in average correlation energies, the fitted equilibrium bond length ($a_{\text{min}}$) for RCI is very similar to that of NNCI (52 MOs): $1.1044$~\AA{} vs. $1.1047$~\AA.
We attribute this deviation to fitting uncertainties (estimated at $2\text{--}3$~m\AA) due to the grid discretization, and the insufficient basis functions which generally overestimate the triple bond length, rather than intrinsic limitations of the CI space determination.

\hy{
Both NNCI and RCI have similar stochastic elements in the iterative subspace expansion procedure, e.g., the random subset sampling of the pool and the network initialization.
To compare the two methods more robustly, we performed five independent runs with different random seeds for each method at five different bond lengths across the \ce{N2} dissociation curve and used Welch's $t$-test to assess whether the mean final energy of RCI is significantly lower than that of NNCI (see Supporting Information for details).
Across the five bond lengths tested, RCI's mean energy is significantly lower than NNCI's in four cases ($p$-value $< 0.05$), with the remaining case still favoring RCI, though not reaching statistical significance.
}

In conclusion, these results demonstrate that RCI is capable of recovering lower correlation energies across the entire \ce{N2} dissociation curve while requiring significantly less computational time.

\subsection{\hy{Basis-Set Scaling of Convergence Efficiency}}
\label{sec:basis-scaling}

\hy{
A natural concern with SCI approaches is whether their efficiency erodes as the basis set grows.
In principle, enlarging the basis introduces a rapidly increasing number of virtual orbitals and, correspondingly, a much larger pool of weakly coupled determinants with small CI coefficients. 
If the method's performance depended on exploiting only the largest coefficients, one might expect a ``diminishing returns'' regime: as the determinant space is enlarged, previously important coefficients could become only weakly connected to the leading configurations, degrading the efficiency of the selection procedure and eroding the smoothness-based compression advantage~\cite{griebel2007sparse,chinnamsetty2018adaptive,anderson2018breaking}.}

\hy{The \ce{BeH2} molecule is a well-studied benchmark system owing to the near-degenerate 2s and 2p orbitals of the beryllium atom~\cite{BeH2_r,BeH2_p1,BeH2_p2}. 
To test the basis-set scaling of RCI, we studied this molecule at its equilibrium geometry (Be--H bond length $1.33408$~\AA~\cite{BeH2_r}) and at a stretched geometry (Be--H bond length $2.50$~\AA), including all electrons across three basis sets of increasing size: 6-31G, cc-pVDZ, and cc-pVTZ, comprising 13, 24, and 58 MOs, respectively. 
This sequence provides a controlled ladder of orbital-space growth, allowing us to explore the effect of basis-set size on convergence behavior.}

\begin{figure*}[htbp]
    \centering
    \includegraphics[width=1\textwidth]{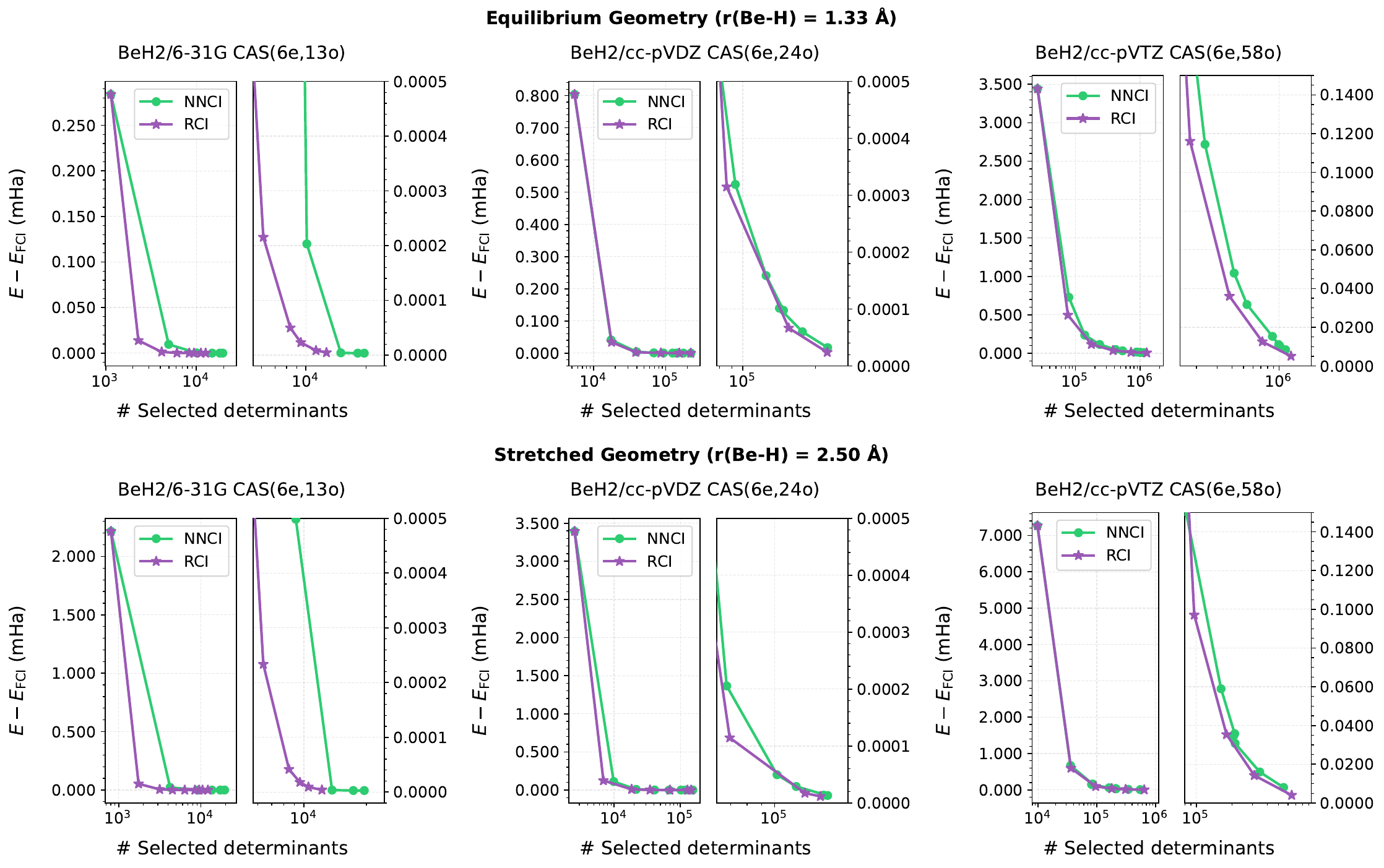}
    \caption{
    \hy{Convergence of RCI and NNCI on \ce{BeH2} at the equilibrium ($1.33408$~\AA) and stretched ($2.50$~\AA) Be--H bond lengths, across three basis sets: 6-31G, cc-pVDZ, and cc-pVTZ.
    The inflation factor $\gamma$ of RCI is set to $1.3$.
    For the equilibrium geometry, the FCI energies are used as reference values, which are $-15.80040751$~Ha, $-15.83643474$~Ha, and $-15.85694113$~Ha for the three basis sets, respectively. 
    For the stretched geometry, the corresponding FCI energies are $-15.62979015$~Ha, $-15.64266648$~Ha, and $-15.65670322$~Ha, respectively.
    }
    }
    \label{fig:beh2-scaling}
\end{figure*}

\hy{
Figure~\ref{fig:beh2-scaling} shows that, at both the equilibrium and stretched geometries, both NNCI and RCI achieve high accuracy using only a small fraction of the full Hilbert space: even for the largest active space considered, CAS(6e,58o), chemical accuracy is reached with approximately $0.01\%$ of the full Hilbert space.
Moreover, RCI consistently shows an advantage over NNCI, requiring fewer determinants to reach a given accuracy across all three basis sets. 
The qualitative shape of the RCI convergence curve is preserved across all three panels: a steep initial descent followed by a gradually decreasing step size as it approaches the FCI limit.
In contrast, NNCI's convergence curves are less stable, especially for larger basis sets, exhibiting irregular step sizes that alternate between small and large jumps during the iterations.
This robustness indicates that the underlying mechanism, namely the rapid convergence of the CI expansion driven by the smoothness of the wavefunction, continues to operate effectively in RCI even as the full determinant space grows by several orders of magnitude.}

\hy{These results provide direct empirical evidence against a basis-set-driven diminishing-returns effect for the systems tested here. 
We acknowledge, however, that in systems with more pronounced ground-state degeneracy or strong symmetry-breaking tendencies, distinct symmetry-broken solutions can give rise to isolated ``islands'' of large coefficients in the Hilbert space that compete for selection during the SCI expansion. 
Extending such scaling studies to these more demanding systems therefore remains an important direction for future work.
}

\subsection{Results on \ce{[Fe2S2(SCH3)4]^{2-}}}

Next, we assess the capability of RCI in treating strongly correlated systems by investigating the synthetic \ce{[Fe2S2(SCH3)4]^{2-}} cluster~\cite{mayerle1975synthetic}.
It features two iron centers bridged by two sulfides, with each iron additionally coordinated by two methylthiolate (\ce{CH3S-}) ligands.
The electronic structure of this system is notoriously difficult to describe, owing to the presence of multiple low-lying states with distinct spin and charge characters, as well as the subtle superexchange interactions mediated by the sulfur ligands~\cite{sharma2014low}.
These complexities render it a stringent benchmark for advanced SCI methodologies.
Our calculations employ an active space of 30 electrons in 20 spatial orbitals, CAS(30e,20o), which includes the Fe $3d$ and S $3p$ orbitals. 
Both sets of orbitals are crucial for capturing the superexchange mechanism accurately.
The corresponding FCI space dimension is $2.40 \times 10^{8}$.
The one- and two-electron integrals in FCIDUMP format were taken from \citeauthor{li2017spin}~\cite{li2017spin}, who reported a density matrix renormalization group (DMRG) benchmark energy of $-116.605609$~Ha for the spin $S=0$ ground state.

\hy{We compare RCI against the following SCI methods: semistochastic HCI (SHCI)~\cite{shci}, Trimmed CI (TrimCI)~\cite{trimci_1,trimci_2}, and two recent Transformer-based generative SCI approaches: GTNN-SCI~\cite{shang} and HAAR-SCI~\cite{haar-sci}.
All energies reported for these comparisons are variational energies without any perturbative correction.
TrimCI operates on a graph whose nodes are Slater determinants and edges correspond to nonzero Hamiltonian couplings. 
Starting from random determinants, it iteratively expands the variational space along strong Hamiltonian couplings and trims away unimportant states, converging to a compact core that accurately approximates the ground state.}
Both GTNN-SCI and HAAR-SCI adopt Transformer decoders to model the conditional probability of each orbital's occupation conditioned on all preceding ones. 
The models are trained by minimizing the MSE between the predicted CI coefficients and the reference coefficients obtained from Hamiltonian diagonalization. 
After training, determinants with large joint probabilities are generated via autoregressive sampling and subsequently used to expand the variational subspace.

As summarized in Table~\ref{tab:method_comparison}, RCI delivers highly competitive and robust performance on this challenging system, achieving an energy error of 1.35~mHa relative to the DMRG reference using only $2.80 \times 10^7$ determinants (approximately 12\% of the full Hilbert space). 
\hy{At comparable determinant counts (approximately $2.80 \times 10^7$), SHCI and HAAR-SCI yield errors of 2.3~mHa and 1.84~mHa, respectively. 
GTNN-SCI and TrimCI only marginally reach the chemical accuracy threshold (1.60~mHa and 1.56~mHa, respectively), whereas RCI attains 1.35~mHa, which is more than 15\% below this threshold.
Furthermore, HAAR-SCI requires $3.33 \times 10^7$ determinants to achieve chemical accuracy (1.43~mHa), whereas RCI attains comparable accuracy (1.35~mHa) using 16\% fewer determinants, underscoring its superior efficiency. 
At a comparable determinant count of approximately $3.50 \times 10^7$, RCI's energy error is 21\% lower than that of TrimCI (1.18~mHa vs. 1.50~mHa).}

In conclusion, the LTR strategy improves determinant selection efficiency in our RCI approach. By reformulating the core selection task as a ranking problem rather than a regression problem, RCI identifies the most important determinants more effectively, thereby yielding a more compact and accurate variational subspace. 
This makes RCI particularly suitable for strongly correlated systems.

\begin{table}[htbp]
    \centering
    \caption{Performance Comparison of Different SCI Methods on the \ce{[Fe2S2(SCH3)4]^{2-}} Cluster$^a$}
    \label{tab:method_comparison}
    \begin{tabular}{l r r}
        \hline\hline
        Method & $N_{\text{det}} \times 10^7$ & $E_{\text{var}} - E_{\text{DMRG}}$ (mHa)\\
        \hline
        SHCI~\cite{shang}                   & $2.80$        & $2.30$  \\
        \hline
        GTNN-SCI~\cite{shang}               & $\sim 2.80$   & $1.60$  \\
        \hline
        \hy{TrimCI}                              & \hy{$2.80$}        & \hy{$1.56\pm 0.00$}  \\
        \hy{TrimCI}                              & \hy{$3.50$}        & \hy{$1.50\pm 0.00$}  \\
        \hline
        \hy{HAAR-SCI}                            & \hy{$2.91$}        & \hy{$1.84$}  \\
        \hy{HAAR-SCI}                            & \hy{$3.33$}        & \hy{$1.43$}  \\
        \hline
        \hy{RCI (ours)}                          & \hy{$2.80$}        & \hy{$1.35\pm 0.01$} \\
        \hy{RCI (ours)}                          & \hy{$3.30$}        & \hy{$1.23\pm 0.01$} \\
        \hy{RCI (ours)}                          & \hy{$3.50$}        & \hy{$1.18\pm 0.01$} \\
        \hline\hline
    \end{tabular}
    
    \vspace{1ex}
    \raggedright
    \footnotesize \hy{$^a$ 
    All energies reported are variational energies.
    $N_\text{det}$ is the number of selected determinants in the variational wavefunction.
    $E_{\text{var}}$ is the variational energy, and $E_\text{DMRG} = -116.605609$~Ha is taken as the reference energy~\cite{li2017spin}.
    For GTNN-SCI, the comparison is only indicative, as the determinant count required to reach the chemical accuracy threshold ($1.60$~mHa) is estimated from the graphical data in the original publication.
    For fairness, the variant of TrimCI without orbital optimization is adopted.
    The mean and standard deviation reported for TrimCI and RCI are computed over five independent runs.
    Detailed results of TrimCI and RCI are provided in the Supporting Information.
    HAAR-SCI results are based on a single run (no variance estimate available) due to prohibitive computational cost.}
\end{table}

\subsection{\hy{Results on \ce{Cr2}}}

\hy{
In this section, we apply RCI to the chromium dimer (\ce{Cr2}), which serves as a popular and challenging correlated system for benchmarking electronic structure methods~\cite{cr2_shci,cr2_dmrg,cr2_chapter}.
We employ a CAS(48e,42o) active space, corresponding to a Hilbert space of approximately $10^{23}$.
The one- and two-electron integrals are obtained from a CASSCF(12e, 12o) calculation using the Ahlrichs' SV basis set at a bond length of $1.5$~\AA~\cite{che2026}, ensuring direct comparability with prior SHCI~\cite{cr2_shci} and DMRG~\cite{cr2_dmrg} results.
Pruned wavefunctions are employed to assess wavefunction compactness and to reduce the perturbative cost.
}

\hy{
The results in Table~\ref{tab:cr2} demonstrate that both SHCI and RCI can reach a comparable variational energy to single-reference CCSDTQ ($-0.430$ Ha), requiring 2,237,828 and 1,467,711 determinants, respectively, which corresponds to a 34\% reduction in the number of determinants for RCI. 
With EN-PT2 correction, the total energies at these determinant counts remain comparable. 
Furthermore, when using a similar determinant count to SHCI (2,114,621 vs.\ 2,237,828), RCI achieves a variational energy more than 2 mHa lower than that of SHCI, and the total energies of RCI and SHCI lie 0.65 mHa and 0.88 mHa above the extrapolated DMRG energy, respectively.
These results indicate that RCI can yield a more compact and accurate wavefunction than SHCI on the \ce{Cr2} system.
}

\begin{table}[htbp]
\centering
\caption{
\hy{Comparison of Variational and Total Energies Obtained by Selected Methods for the \ce{Cr2} Dimer.$^a$ }
}
\label{tab:cr2}
    \begin{tabular}{lrrr}
        \hline\hline
        Method & $N_{\text{det}}$ & $E_{\text{var}}$ (Ha) & $E_\text{tot}$ (Ha) \\
        \hline
        CCSD(T)~\cite{cr2_dmrg}     & -      & -          & $-0.422229$ \\
        CCSDTQ~\cite{cr2_dmrg}      & -      & -          & $-0.430244$ \\
        \hline
        SHCI~\cite{cr2_shci}        & $787,919$     & $-0.422163$ & $-0.443463$ \\
        SHCI~\cite{cr2_shci}        & $2,237,828$   & $-0.430093$ & $-0.443908$ \\
        \hline
        RCI (ours)                  & $972,574$     & $-0.427875$ & $-0.443579$ \\
        RCI (ours)                  & $1,467,711$   & $-0.430692$ & $-0.443977$ \\
        RCI (ours)                  & $2,114,621$   & $-0.432791$ & $-0.444130$ \\
        \hline
        DMRG~\cite{cr2_dmrg}        & -             & $-0.443334$ & $-0.444784$ \\
        \hline\hline
    \end{tabular}

    \vspace{1ex}
    \raggedright
    \footnotesize \hy{$^a$ 
    All energies are reported as $E + 2086$ Ha.
    $N_\text{det}$ denotes the number of selected determinants.
    $E_{\text{var}}$ is the variational energy, and $E_\text{tot}$ is obtained by adding a second-order Epstein--Nesbet perturbation theory (PT2) correction to $E_{\text{var}}$.}
\end{table}

\subsection{Ablation Study}\label{sec:ablation}

Finally, we performed comprehensive ablation studies to analyze the individual contributions of the core algorithmic components within the RCI framework. 
Specifically, our investigation focuses on two primary aspects: first, the synergy between the model architecture and the training objective; and second, the impact of the data sampling strategy on training efficiency.

\subsubsection{Analysis of Model Architecture and Training Objective}

The proposed RCI distinguishes itself from prior ML-supported SCI methods (e.g., NNCI) in two primary aspects: the model architecture (Transformer vs. convolutional neural network (CNN)/MLP) and the training objective (LTR vs. classification/regression).
To disentangle the contributions of these components, we evaluated the performance of four distinct combinations.
In this context, the representative NNCI method serves as the CNN+Classification baseline, and RCI corresponds to the Transformer+LTR configuration.
Consequently, we constructed two intermediate variants: Transformer+Classification (replacing only the architecture) and CNN+LTR (replacing only the training objective).
To ensure a fair comparison of model architecture and training objective, all variants use the same dynamic determinant selection strategy ($\gamma=1.5$) during the SCI iterations.

\begin{figure*}[ht]
\centering 
\includegraphics[width=0.9\textwidth]{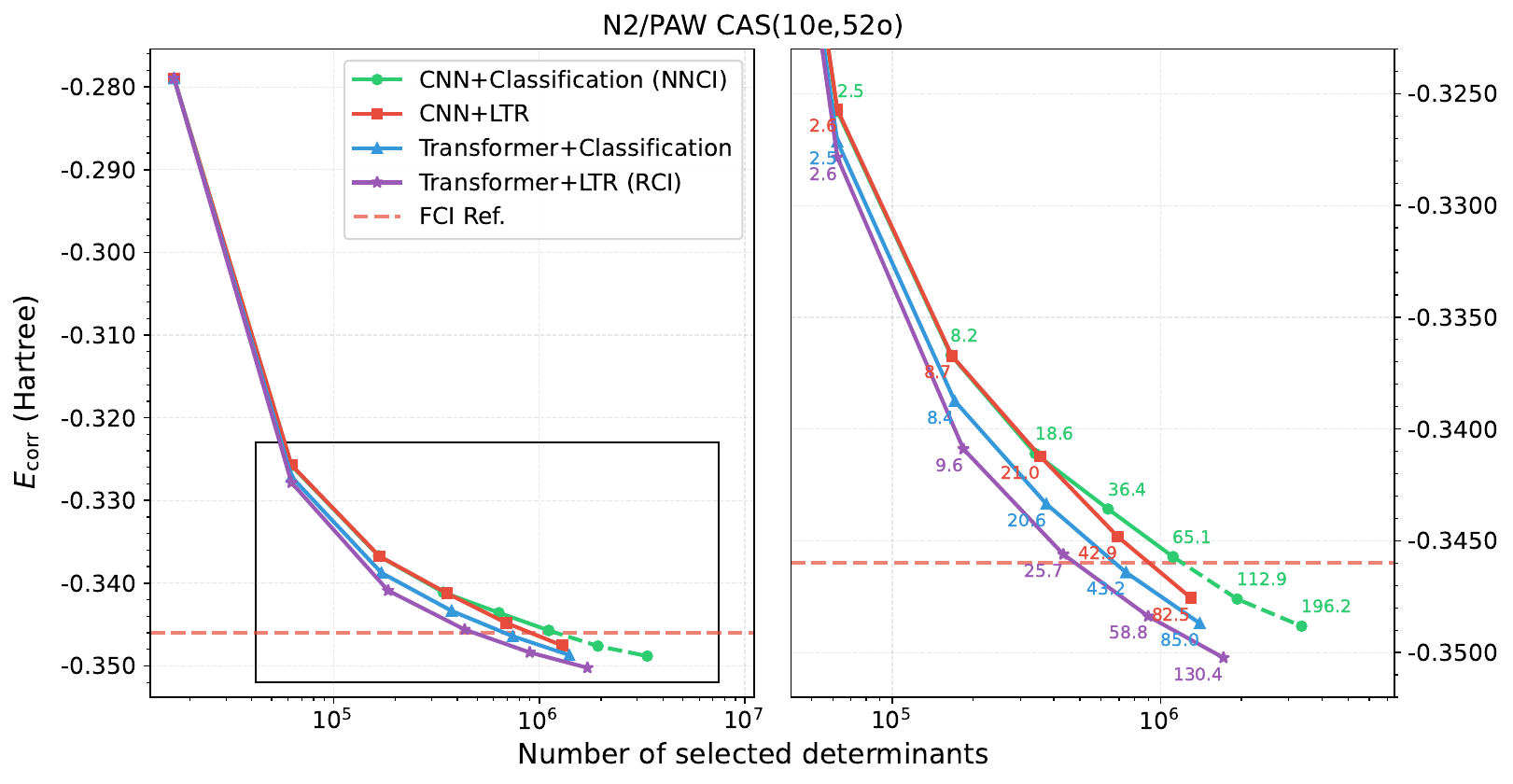} 
\caption{
Ablation study on the effects of model architecture and training objective. 
The recovered correlation energy for \ce{N2} (plane-wave basis, CAS(10e,52o)) is plotted against the number of selected determinants, where a steeper descent indicates more efficient determinant selection.
Selected wall times are annotated near the corresponding data points, in hours.
The reference FCI energy (FCI Ref.) is taken from previously reported \ce{N2} calculations with 34 MOs~\cite{fci_N2}.
Four model variants are compared: CNN+Classification (NNCI), CNN+Learning to Rank (LTR), Transformer+Classification, and the proposed Transformer+LTR (RCI). 
All methods are evaluated for six iterations, except for the CNN+Classification (NNCI) baseline, which is extended by two additional iterations (dashed line) to provide a reference at comparable energy levels.
Replacing either the CNN encoder with a Transformer or the classification objective with LTR consistently accelerates convergence, while their combination in RCI yields the best overall performance. 
}
\label{fig:ablation}
\end{figure*}

As illustrated for \ce{N2} in the plane-wave basis (Figure~\ref{fig:ablation}), replacing the CNN+Classification (NNCI) baseline components with either the Transformer architecture or the LTR objective consistently accelerates convergence, with their combination (RCI) achieving the most rapid energy descent. 
Specifically, the Transformer+Classification, CNN+LTR and RCI models all successfully recover correlation energies below the FCI reference~\cite{fci_N2} within just six iterations. 
To provide a point of reference, the baseline method requires seven to eight iterations to reach comparable energy levels, and even then, the proposed RCI method delivers a final energy more than $1$ mHa lower. 
Interestingly, although all methods employ the identical dynamic determinant selection strategy ($\gamma=1.5$), this systematic improvement in energy convergence is directly mirrored by the sizes of their final variational spaces, which establish a clear hierarchy: Transformer+LTR (RCI) $>$ Transformer+Classification $>$ CNN+LTR $>$ CNN+Classification (NNCI). 
This divergence in subspace size stems from the inherent differences in model predictive accuracy. 
Models with lower fidelity identify a smaller proportion of genuinely important determinants during the selection phase (see Supporting Information for details). 
Consequently, fewer determinants survive the subsequent pruning step. Over successive iterations, this inefficient expansion restricts the growth of the variational space, leaving less accurate methods with a noticeably smaller cumulative subspace at the same iteration.

In conclusion, this ablation study demonstrates two key points.
First, the Transformer architecture is significantly more efficient than CNN at learning the complex features of determinants, and the LTR objective captures relative importance more effectively than binary classification; both independently accelerate convergence.
Second, there is a strong synergy between the two: the LTR objective inherently requires the model to discern subtle, fine-grained relative differences to establish a correct ranking, and the Transformer provides the necessary expressive capacity to fully capture these intricate features, thereby overcoming the representational bottlenecks of simpler architectures.

\subsubsection{Analysis of Active Pair Sampling Strategy}
\begin{figure*}[ht]
\centering 
\includegraphics[width=0.9\textwidth]{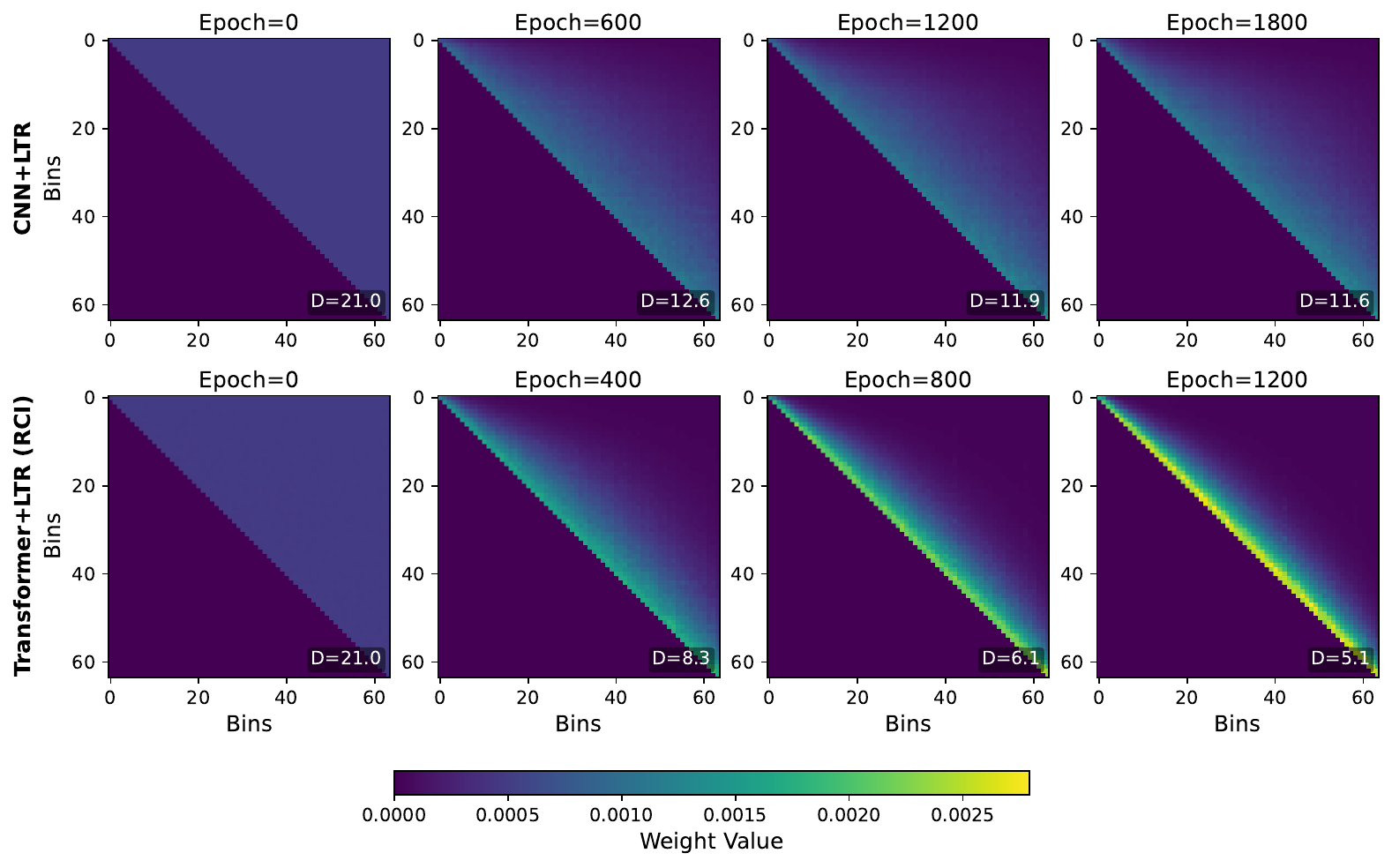} 
\caption{
Analysis of the active pair sampling strategy during pairwise ranking training.
The heatmaps illustrate the evolution of the sampling weight matrix $W$ across different training epochs for the CNN+LTR baseline (top) and the Transformer-based RCI model (bottom). 
Each entry $W_{ij}$ denotes the sampling probability associated with the bin pair $(i,j)$ for drawing determinant pairs.
Because determinants are sorted in descending order of coefficient magnitude, only bin pairs with $i \leq j$ form valid ordered pairs, making $W$ an upper triangular matrix.
As training proceeds, the weight distribution progressively concentrates toward the diagonal. 
This shift indicates that the model, having mastered the easy distant bin pairs with large coefficient differences, dynamically reallocates its focus toward hard proximal bin pairs with similar magnitudes. 
The faster and sharper diagonal concentration in RCI reflects more efficient hard-pair mining and superior training dynamics. 
}
\label{fig:weight_matrix}
\end{figure*}

To validate the effectiveness of the proposed active pair sampling strategy, we analyze the evolution of the sampling weight matrix $W$ throughout training, as illustrated in Figure~\ref{fig:weight_matrix}.

Initially, the sampling weights $W$ are broadly distributed, reflecting the model's inability to distinguish pairs of varying difficulties. As training progresses, the mass of $W$ progressively concentrates along the diagonal. 
This structural shift---quantitatively captured by a steadily decreasing weighted distance from the diagonal, $D = \sum_{i,j} W_{ij} |i - j|$---indicates that easier distant bin pairs are mastered and assigned lower probabilities. 
Consequently, training attention dynamically shifts toward the more challenging proximal bin pairs, confirming that the active sampling mechanism effectively mines hard negatives.

Furthermore, a quantitative comparison between RCI and the CNN+LTR baseline shows that the weight matrix of RCI converges to a diagonally concentrated pattern both faster and more distinctly. 
Specifically, RCI completes training by $\sim 1200$ epochs, achieving a highly concentrated diagonal distribution with a final weighted distance of $D = 5.1$. 
In contrast, the CNN+LTR baseline requires $\sim 1800$ epochs to converge and still exhibits a noticeably more dispersed weight distribution, yielding a significantly higher final distance of $D = 11.6$.

Taken together, these observations demonstrate that the active pair sampling strategy improves training efficiency by continuously exposing the model to the most informative determinant pairs. 
Moreover, the Transformer-based architecture, owing to its stronger representational capacity and long-range attention mechanism, can resolve easy pairwise distinctions more rapidly, thereby entering the hard-pair regime earlier and maintaining a more informative gradient signal throughout training. 
This, in turn, accelerates convergence on fine-grained coefficient ordering and leads to better overall performance, as also evidenced in Figure~\ref{fig:ablation}.

\section{Discussion}

In this work, we introduced RCI, a novel ML-driven framework for selected CI that reframes determinant selection as an LTR problem. 
Traditional supervised approaches in this domain typically rely on pointwise regression or classification, inherently suffering from an objective-loss mismatch. 
By introducing the LTR paradigm to SCI, RCI employs a pairwise loss that optimizes a much tighter proxy for the relative ordering of determinants, aligning the training objective more closely with the fundamental goal of SCI:
prioritizing the most variationally significant configurations. 
To fully leverage this ranking objective, we employed a Transformer-based architecture, whose self-attention mechanism effectively captures the complex, long-range orbital dependencies that simpler models struggle to represent. 
Furthermore, to enhance sample efficiency and ranking fidelity, we incorporated an active pair sampling strategy with hard negative mining, dynamically focusing the training process on the most challenging determinant pairs.

Across benchmark tests utilizing plane-wave basis sets, RCI consistently achieved lower correlation energies at a fraction of the computational cost. 
For instance, reaching \hy{FCI reference energies} in \ce{N2} and \ce{CO} required only $\sim$55\% of the determinant count and less than 46\% of the wall time compared to the classification-based NNCI baseline~\cite{nnci}.
Similar efficiency gains were observed with Gaussian basis sets, where RCI reduced the computational time to 60.3\%--76.9\% of the NNCI cost across \ce{N2}, \ce{C2}, \ce{H2O}, and \ce{NH3} while successfully converging to the exact FCI limits. 
The robustness of our approach was further validated on the \ce{N2} dissociation curve; RCI achieved correlation energies that were, on average, $0.72$~mHa lower than the NNCI results, consuming only 71.5\% of the wall time. 
\hy{To assess whether this efficiency advantage persists as the orbital space grows, we further examined RCI's convergence behavior on \ce{BeH2} across three basis sets of increasing size (6-31G, cc-pVDZ, and cc-pVTZ); RCI consistently required fewer determinants than NNCI to reach a given accuracy across all three basis sets, with no evidence of a basis-set-driven diminishing-returns effect even for the largest active space tested.}
Additionally, on the highly challenging \ce{[Fe2S2(SCH3)4]^{2-}} cluster, RCI reached chemical accuracy with an energy error of $1.35$~mHa relative to the DMRG benchmark using only $\sim$12\% of the full FCI space. 
This represents a more than 15\% improvement in accuracy over recent Transformer-based, regression-driven SCI methods~\cite{shang,haar-sci}.
\hy{Beyond variational-only comparisons, we additionally benchmarked RCI on the substantially larger \ce{Cr2} dimer; at a matched determinant count, RCI achieved both a lower variational energy and a lower PT2-corrected total energy than SHCI, confirming that RCI's advantage over conventional SCI methods holds even when perturbative corrections are included.}
Finally, ablation studies confirmed that the synergy between the Transformer architecture and the LTR objective is crucial for maximizing determinant selection efficiency in RCI. 
Moreover, the active pair sampling strategy proved instrumental in accelerating training by continuously exposing the model to maximally informative determinant pairs.

While these results are promising, we acknowledge certain limitations of the current study and outline avenues for future research. 
The field of ML-supported SCI is evolving rapidly, and a variety of supervised-learning frameworks based on Hamiltonian-guided, adaptive subspace expansion have recently emerged. 
However, these methods predominantly rely on conventional regression or classification objectives. 
The core objective of this study is to investigate whether the LTR paradigm offers a more effective alternative, thereby providing a fresh perspective to the rapidly evolving field of ML-supported SCI. 
To this end, we primarily selected the open-source classification-based NNCI as our baseline for systematic comparison of determinant counts and wall times, replacing its classification objective with LTR while keeping the Hamiltonian-guided iterative SCI procedure largely intact. 
Given the diverse complexities and varying degrees of open-source availability among existing methods, a systematic benchmarking across all contemporary variants lies beyond the current scope of this work. 
Nevertheless, this proposed pairwise LTR model provides a lightweight, modular plugin that can be seamlessly incorporated into other supervised-learning frameworks, regardless of model architecture (MLP, CNN, or Transformer) or training objective (regression or classification). 
A comprehensive benchmarking of such integrations remains an important direction for future work and will be essential for assessing the broader applicability of the proposed approach.

\begin{acknowledgement}

The authors acknowledge financial supports from the Hong Kong Research Grants Council through General Research Funds (17307323, 17305724),
and the use of generative AI tools for language refinement to enhance the readability of this manuscript. 
W.N. thanks Yiran Wang, Zheng Che, and Hao Zhang for helpful discussions.

\end{acknowledgement}

\begin{suppinfo}

The Supporting Information is available free of charge at

\begin{quote}
Hyperparameters and configuration of the RCI framework;
\hy{
TrimCI variational energies on the \ce{[Fe2S2(SCH3)4]^{2-}} cluster;
comparison of loss functions and model architectures for RCI;
analysis of the dynamic selection strategy across RCI iterations;
statistical comparison of RCI and NNCI on the \ce{N2} dissociation curve across five independent runs;
five independent runs of RCI on the \ce{[Fe2S2(SCH3)4]^{2-}} cluster;
}
and fraction of important determinants retained after the pruning process across iterations for the evaluated model variants in the ablation study.
\end{quote}

\end{suppinfo}


\section{Data and Code Availability}
All code and data developed in this study have been made publicly available on GitHub at \url{https://github.com/wan-nie/RCI.git}.

\section{Competing Interests} 
The authors declare no competing interests.

\bibliography{manuscript}

\end{document}


\clearpage
\section{Supporting Tables}

\begin{table}[htbp]
\centering
\caption{Hyperparameters and configuration of the RCI framework.$^a$}
\label{tab:hyperparameters}
\begin{threeparttable}
    \begin{tabular}{
        p{0.20\textwidth}
        p{0.10\textwidth}
        p{0.60\textwidth}
        p{0.10\textwidth}}
    \toprule
    \textbf{Stage} & \textbf{Notation} & \textbf{Description} & \textbf{Value} \\
    \midrule
    Sampling
      & $p^{\text{rand},t}$ & Fraction of $\{\phi^{\text{pool},t}\}$ to select $\{\phi^{\text{rand},t}\}$ 
                                                                                    & $\frac{50}{\sqrt{n}} \times \frac{4}{3}$ \\
    \midrule
    Model Architecture
      & $d$                & Spin orbital embedding dimension                       & $256$  \\
      & $d_{\text{model}}$ & Transformer encoder model dimension                    & $256$  \\
      & $N_{\text{head}}$  & Number of attention heads                              & $2$    \\
      & $d_{\text{ff}}$    & Transformer encoder feedforward dimension              & $512$  \\
    \midrule
    Training Procedure
      & $N_{\text{pair}}$   & Determinant pairs batch size                          & $\frac{n(n-1)}{1000}$ \\
      & $N_{\text{bins}}$   & Number of bins                                        & $64$   \\
      & $N_{\text{probe}}$  & Number of determinant pairs sampled for each bin pair & $N_{\text{pair}}/N_{\text{bins}}^{2}$ \\
      & $N_{\text{update}}$ & Weight update interval (epochs)                       & $10$   \\
      & $\tau$              & Temperature exponent for loss sharpening              & $1.5$  \\
      & $\varepsilon$       & Minimum sampling probability for bin pairs            & $0.05$ \\
      & $\lambda$           & EMA decay for sampling weight matrix                  & $0.2$  \\
      & $\epsilon$          & Tolerance for Spearman $\rho$                         & $10^{-6}$ \\
      & $N_{\text{epoch}}$  & Maximum training epochs                               & $5000$ \\
      & $N_{\text{eval}}$   & Evaluation interval (epochs)                          & $10$   \\
      & $\eta$              & Learning rate                                         & $10^{-3}$ \\
      & $N_{\text{estop}}$  & Early stopping patience                               & $5$    \\
    \midrule
    Selecting
      & $p^{\text{impt},t}$
      & Fraction of $\{\phi^{\text{rand},t}\}$ to define the coefficient threshold for selecting $\{\phi^{\text{rand\_impt},t}\}$ and pruning $|\Psi^{\text{gs},t}\rangle$
      & $\frac{50}{\sqrt{n}}$ \\
      & $\gamma$            & Inflation factor for dynamic selection in the pool    & $1.5$    \\
      & $p^{\text{cap}}$    & Cap on the dynamic ranking selection ratio            & $0.2$    \\
    \bottomrule
    \end{tabular}
\begin{tablenotes}[flushleft]
  \footnotesize
  \item $^a$ $n$ denotes the size of the candidate pool $\{\phi^{\text{pool},t}\}$. 
  $N_{\text{pair}}$ and $N_{\text{probe}}$ are capped at $10^8$ and $100$, respectively, to limit memory and computational overhead.
  $p^{\text{rand},t}$ and $p^{\text{impt},t}$ follow the settings of NNCI~\cite{nnci}.
\end{tablenotes}
\end{threeparttable}
\end{table}

\clearpage
\begin{table}[htbp]
\centering
\caption{TrimCI variational energies on the \ce{[Fe2S2(SCH3)4]^{2-}} cluster with different number of determinants.$^a$}
\label{tab:dmrg_results}
\begin{threeparttable}
    \begin{tabular}{ccc}
        \toprule
        \textbf{Number of determinants} & \textbf{Energy (Ha)} & \textbf{Energy error (mHa)} \\
        \midrule
        \multirow{5}{*}{$28,000,000$}
            & $-116.60404948$ & $1.55961576$ \\
            & $-116.60404938$ & $1.55971766$ \\
            & $-116.60404943$ & $1.55966945$ \\
            & $-116.60404941$ & $1.55968954$ \\
            & $-116.60404932$ & $1.55977695$ \\
        \cmidrule(lr){2-3}
            &  & $1.55969387 \pm 0.00005955$ \\
        \midrule
        \multirow{5}{*}{$35,000,000$}
            & $-116.60410605$ & $1.50304725$ \\
            & $-116.60410606$ & $1.50303856$ \\
            & $-116.60410611$ & $1.50299477$ \\
            & $-116.60410603$ & $1.50307200$ \\
            & $-116.60410603$ & $1.50307454$ \\
        \cmidrule(lr){2-3}
            &  & $1.50304542 \pm 0.00003229$ \\
        \bottomrule
    \end{tabular}
    \begin{tablenotes}[flushleft]
      \footnotesize
      \item $^a$ Energy errors are relative to the DMRG energy of $-116.605609$~Ha~\cite{li2017spin}.
      Five independent runs are recorded, with the mean and standard deviation provided.
    \end{tablenotes}
\end{threeparttable}
\end{table}


\clearpage
\section{Supporting Figures}

\begin{figure}[ht]
\centering 
\includegraphics[width=0.9\textwidth]{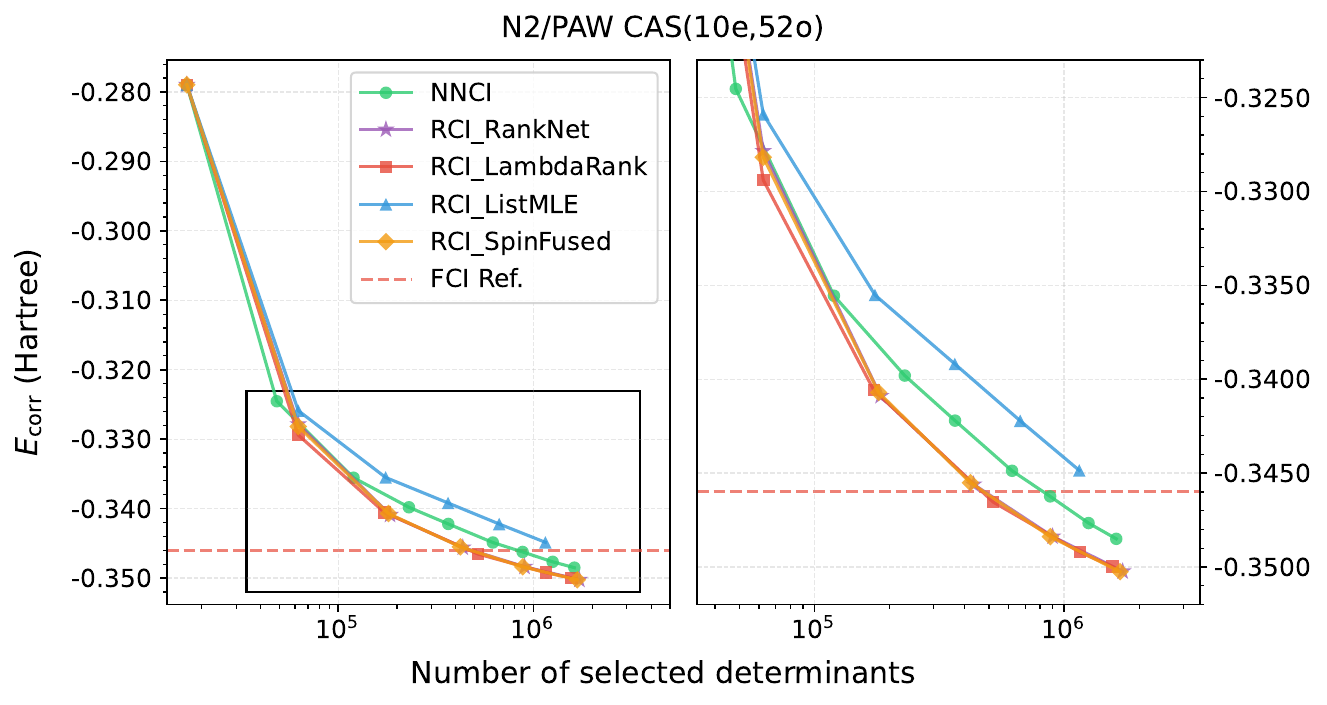} 
\caption{
Comparison of loss functions and model architectures for RCI.
The recovered correlation energy for \ce{N2} (plane-wave basis, CAS(10e,52o)) is plotted against the number of selected determinants.
The reference FCI energy (FCI Ref.) is taken from previously reported \ce{N2} calculations with 34 MOs~\cite{fci_N2}.
Four RCI variants are compared: RCI trained with RankNet (default), LambdaRank, or ListMLE loss, and RCI with a fused-spin-channel architecture, all evaluated over six iterations.
The iteration trajectory of NNCI is included for reference.
ListMLE yields the worst performance across all iterations.
RankNet and the fused-spin-channel variant follow a comparable convergence trajectory throughout.
LambdaRank converges faster in early iterations but exhibits instability at the final iteration, ultimately achieving a similar energy to RankNet and the fused-spin-channel variant.
}
\label{fig:loss_spin}
\end{figure}

\clearpage
\begin{figure}[ht]
\centering 
\includegraphics[width=1.0\textwidth]{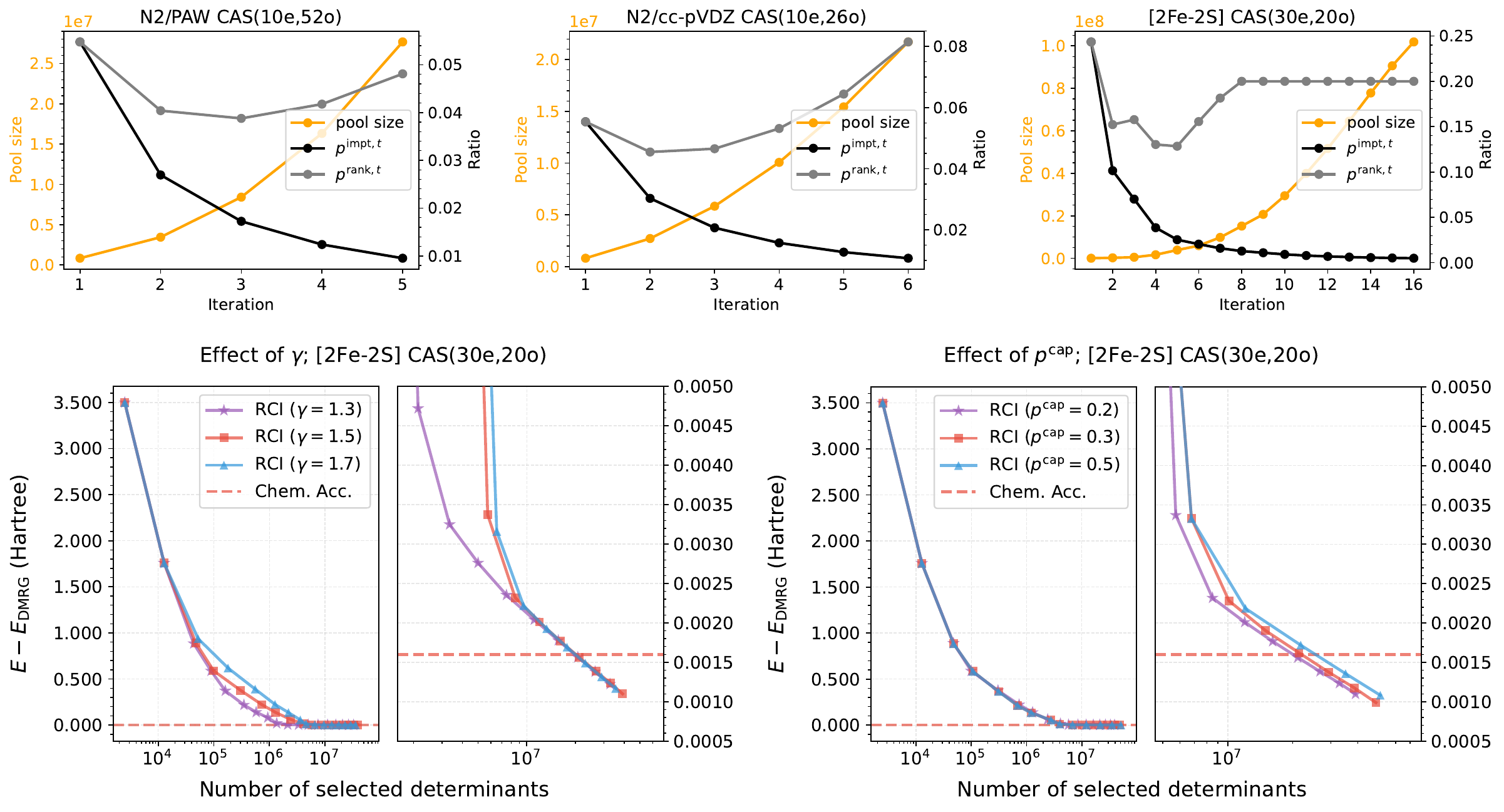} 
\caption{
Analysis of the dynamic selection strategy across RCI iterations.
\textbf{Top:} evolution of the pool size, $p^{\text{impt},t}$, and $p^{\text{rank},t}$ over the RCI iterations for \ce{N2} (plane-wave and cc-pVDZ basis sets) and the \ce{[Fe2S2(SCH3)4]^{2-}} cluster ([2Fe--2S]), with the cap ratio $p^{\text{cap}}$ set to $0.2$.
The pool size is read on the left $y$-axis, and the ratio values on the right $y$-axis.
The pool size expands roughly exponentially with the iteration index, driving a corresponding decrease in $p^{\text{impt},t}$, while $p^{\text{rank},t}$ rises in tandem until it saturates at the cap ratio $p^{\text{cap}}$.
\ce{N2} converges within a few iterations and never reaches the cap, whereas the [2Fe--2S] cluster requires more iterations and reaches the cap.
Note that the initial value of $p^{\text{rank},t}$ for [2Fe--2S] exceeds $0.2$; this reflects a safeguard built into the implementation, whereby the cap is intended solely to suppress runaway inflation of $p^{\text{rank},t}$, rather than to penalize the baseline $p^{\text{impt},t}$ when it already exceeds the cap.
\textbf{Bottom:} sensitivity analysis of the inflation factor $\gamma$ and the cap ratio $p^{\text{cap}}$ for the [2Fe--2S] cluster.
The energy error relative to the DMRG reference~\cite{li2017spin} ($-116.605609$~Ha) is shown as a function of the number of selected determinants.
The inflation factor $\gamma$ predominantly governs the compactness of the wavefunction in the early iterations, before $p^{\text{cap}}$ is reached, whereas $p^{\text{cap}}$ dominates the later iterations, once it becomes active.
A smaller $\gamma$ enables more selective identification of important determinants, yielding a more compact wavefunction; similarly, a smaller $p^{\text{cap}}$ enforces greater compactness in the later iterations.
}
\label{fig:Fe2S2_select}
\end{figure}

\clearpage
\begin{figure}[ht]
\centering 
\includegraphics[width=1\textwidth]{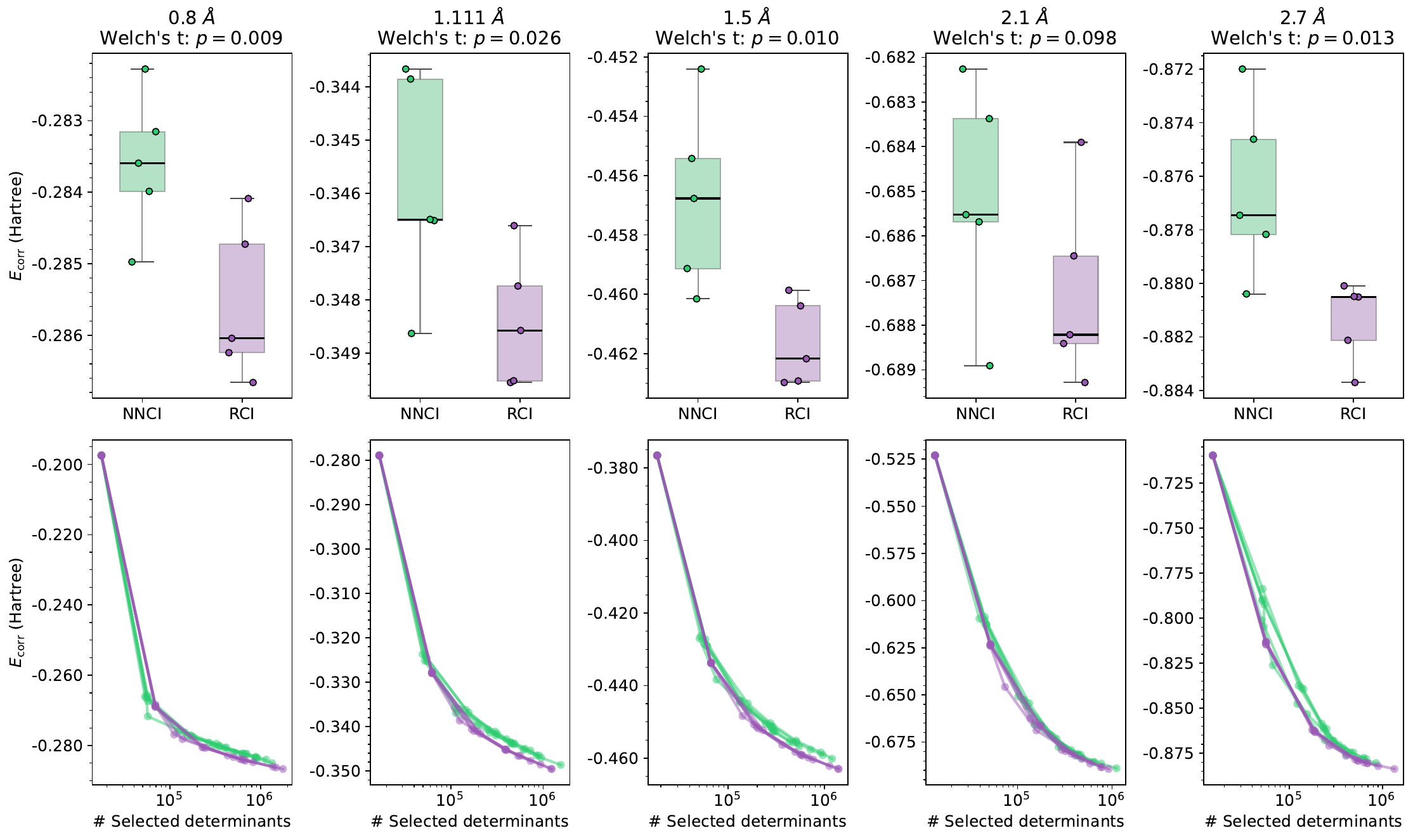} 
\caption{
Comparison of NNCI (green) and RCI (purple) on the \ce{N2} molecule (plane-wave basis, CAS(10e,52o)) across five bond lengths: $0.8$, $1.111$, $1.5$, $2.1$, and $2.7$~\AA, based on five independent runs per method at each bond length.
\textbf{Top:} Box plots of the final variational energies for NNCI and RCI, with Welch's $t$-test used to assess whether RCI's mean energy is significantly lower than NNCI's. 
At all bond lengths, RCI achieves a significantly lower mean energy than NNCI ($p$-value $< 0.05$), except at $2.1$~\AA, where the mean is still approximately $3$~mHa lower despite not reaching significance.
\textbf{Bottom:} Convergence of correlation energy with respect to the number of selected determinants for NNCI and RCI at the corresponding bond lengths.
All five independent runs are plotted together in the same panel to visualize the spread of convergence trajectories.
RCI's convergence curves generally lie below those of NNCI, indicating that RCI recovers a comparable amount of correlation energy using significantly fewer determinants.
}
\label{fig:rev1_N2_PEC}
\end{figure}

\clearpage
\begin{figure}[ht]
\centering 
\includegraphics[width=0.9\textwidth]{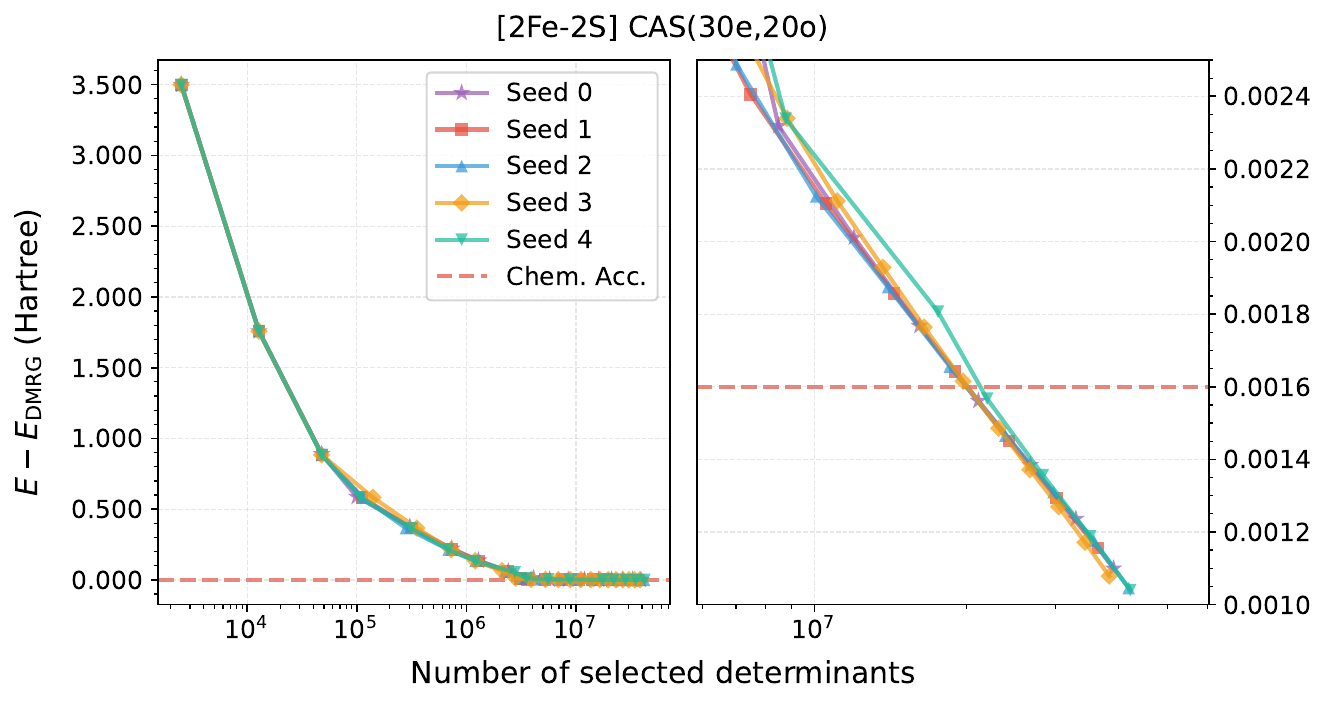} 
\caption{
Five independent runs of RCI on the \ce{[Fe2S2(SCH3)4]^{2-}} cluster.
The energy error relative to the DMRG reference~\cite{li2017spin} ($-116.605609$~Ha) is plotted against the number of selected determinants.
The energy error exhibits linear convergence with respect to $\log_{10}(N_\text{det})$ beyond $10^7$ determinants.
For each run, the energy is estimated at three reference points 
($2.8\times10^7$, $3.3\times10^7$, and $3.5\times10^7$ determinants) 
by linear interpolation between the two iterations straddling each target. 
The mean and standard deviation over five independent runs are calculated separately for each reference point.
}
\end{figure}

\clearpage
\begin{figure}[ht]
\centering 
\includegraphics[width=0.8\textwidth]{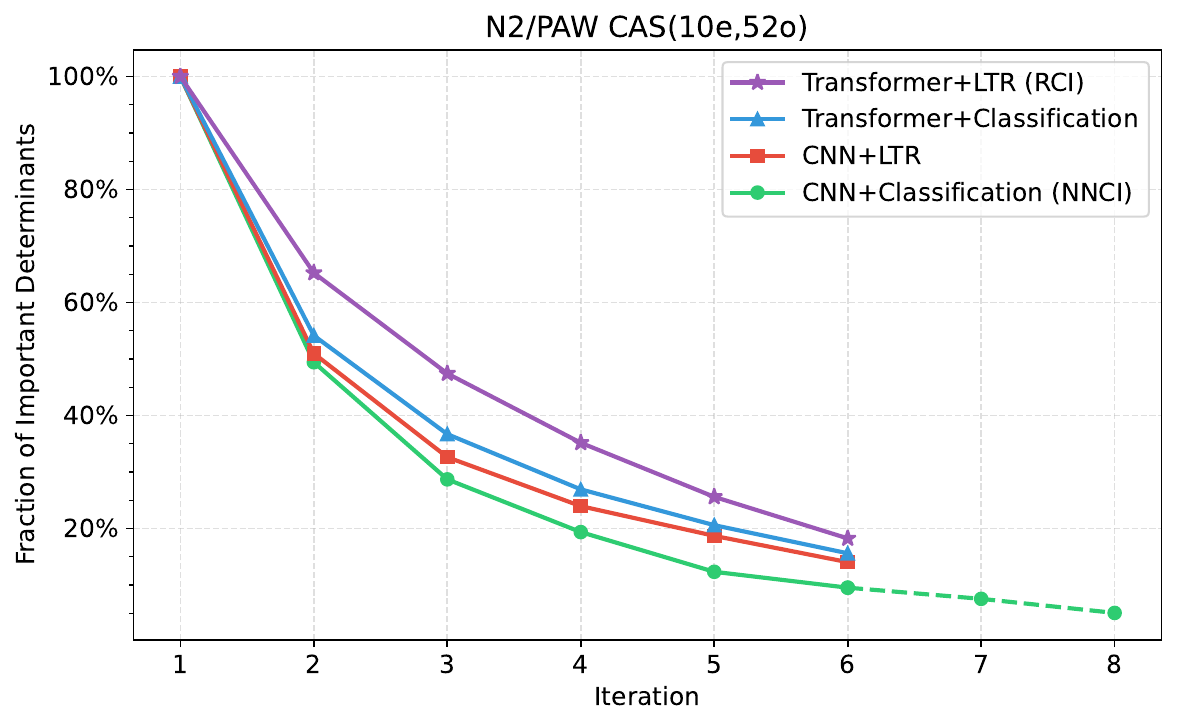} 
\caption{
The fraction of important determinants, retained after the pruning process, in the iterations of the four model variants.
All variants use the same dynamic determinant selection strategy for comparison.
The first iteration corresponds to a non-Machine Learning initialization step, where only an extension operator is applied to generate new determinants without any pruning~\cite{nnci}.
All methods are evaluated for six iterations, except for the CNN+Classification (NNCI) baseline, which is extended by two additional iterations (dashed line) to provide a reference at comparable energy levels.
}
\end{figure}

\clearpage
\bibliography{manuscript}